# A Theory-driven Interpretation and Elaboration of Verification and Validation

Hanumanthrao Kannan[1] and Alejandro Salado[2]
[1]The University of Alabama in Huntsville
[2]The University of Arizona

*Abstract*

This paper presents a formal theory of verification and validation (V&V) within systems engineering, grounded in the axiom that V&V are fundamentally knowledge-building activities. Using dynamic epistemic modal logic, we develop precise definitions of verification and validation, articulating their roles in confirming and contextualizing knowledge about systems. The theory formalizes the interplay between epistemic states, evidence, and reasoning processes, allowing for the derivation of theorems that clarify the conceptual underpinnings of V&V. By providing a formal foundation, this work addresses ambiguities in traditional V&V practices, offering a structured framework to enhance precision and consistency in systems engineering methodologies. The insights gained have implications for both academic research and practical applications, fostering a deeper understanding of V&V as critical components of engineering knowledge generation.

## I. Introduction

In this paper, we take a foundational look at verification and validation (V&V) and try to establish a meaningful theoretical framework for them. What is the meaning of verification? What is the meaning of validation? How do they relate to requirements and needs? Are verification and validation activities mutually exclusive, do they overlap, is one a subset of the other? Certainly, past work has addressed some of these questions. For example, verification is often presented as answering the question of *are you building the system right* and validation as *are you building the right system* [1]. While intuitive, some authors have considered them insufficient and have explored them in more detail, concluding that one must differentiate between kinds of verification (e.g., requirements verification versus system verification) and between kinds of validation (e.g., requirements validation vs system validation) [2, 3]. Yet, these taxonomies lack an underlying and internally consistent theoretical framework that offers constructs to enable formal reasoning. As a result, there is no formal manner to assess the adequacy of the concepts, their correctness, internal consistency, and soundness, leaving us only with our intuition of the problem. Among others, this can lead to confusion, disagreement, and misuse of terms because they do not enable testing the truthfulness of its statements, as has been shown for example in problem formulation when articulating concepts associated with needs and requirements [4, 5]. Recent work has attempted to characterize validation in Systems Engineering using propositional logic, but these efforts remain preliminary and do not yet offer a comprehensive theoretical foundation [6].

Past work that has attempted to take a more formal approach towards conceptualizing V&V has focused on formally modeling verification strategies [7-10], using math to design verification strategies [11-24], or developing insights about the relationship between verification models and requirements [25]. This work used some axiomatic framework to interpret what verification and/or validation were, but the consistency and soundness of the assumptions were not formally tested. This paper helps in bridging that gap.

The focus of this paper is to develop a formal theoretical framework for verification and validation in systems engineering, grounded in the fundamental observation that V&V are knowledge-building activities. While previous work has offered useful but informal definitions and taxonomies for V&V, the field has lacked an internally consistent mathematical foundation that enables formal reasoning about these crucial engineering activities. Using Dynamic Epistemic Modal Logic (DEML) [26, 27], we construct precise definitions of V&V that capture how engineers build knowledge about systems through V&V activities. This formal framework allows us to prove theorems about the relationships between different V&V activities, the role of evidence in building beliefs about system properties, and the conditions under which V&V can overlap or must remain distinct. By providing this theoretical foundation, we resolve longstanding ambiguities in V&V practice and offer systems engineers a structured way to reason about their V&V strategies.

The paper is organized as follows. First, we present a conceptual framework in narrative form to identify the key elements at play in V&V as well as their relationships. Then, we establish and justify the axioms of the proposed theory and follow them with the theoretical framework in the form of formal definitions and some resulting theorems. These are accompanied by discussions relating theory to practice as a way to validate the theoretical constructs. Finally, we suggest research questions that could be explored using the theory presented in the paper.

## II. Framework

As a starting point, we suggest that verification and validation refer to the discovery of how solutions relate to problems. Therefore, we start by identifying the different classes of problem spaces and of solution spaces that are relevant to systems engineering and identify the different relationships between them. These are shown in Figure 1, where the top layer contains elements related to problem spaces and the second layer contains elements related to solution spaces.

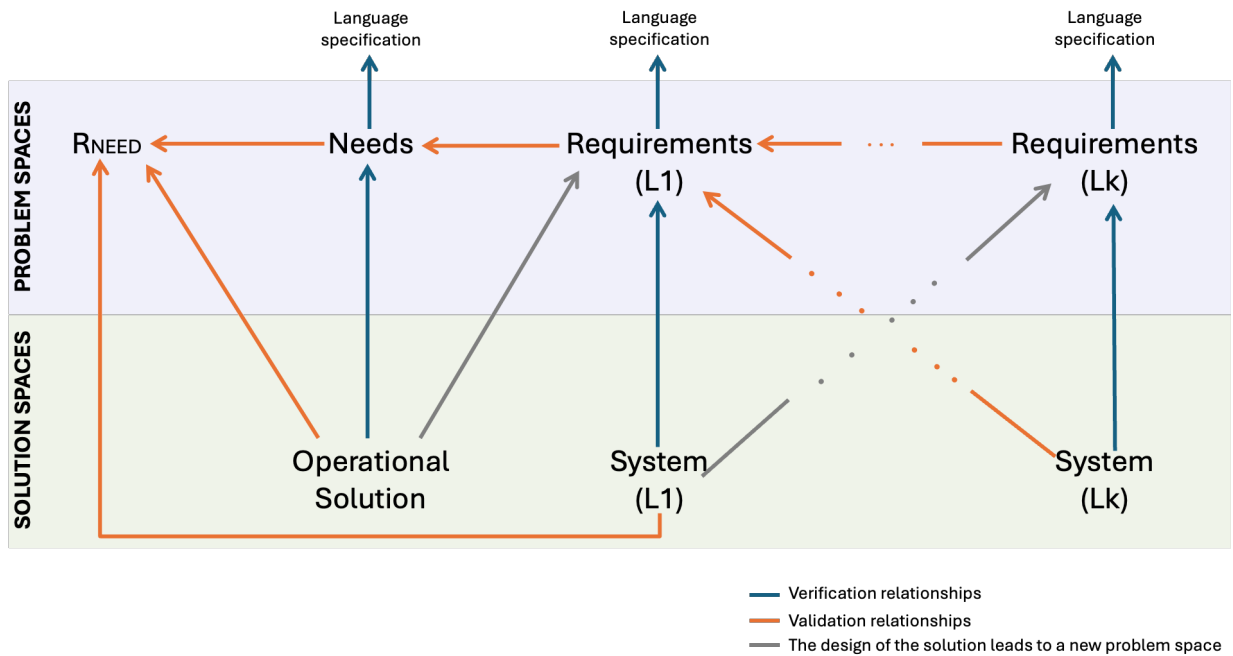

*Figure 1. Conceptual framework*

The concepts related to problems spaces are adopted from [5]. We distinguish two classes of problems, one of desired outcomes and one of desired functions, which are mutually exclusive and considered collectively exhaustive for the practice of systems engineering. We use the term *needs* (or *stakeholder needs*) to refer to the first class and the term *requirements* (or *system requirements*) to refer to the second class. We also distinguish problems from problem spaces. A problem is a desired situational change, whereas a problem space is a representation of such a problem. In other words, where anyone could have a problem in their minds, their articulation is a problem space. This is shown in Figure 1 by differentiating $R_{NEED}$, which refers to the problem that a stakeholder or set of stakeholders have (a desired situational change) from *Needs*, which represents a model of $R_{NEED}$. Finally, we consider that, while problem spaces of outcomes cannot be decomposed but only refined, problem spaces of functions can be decomposed in smaller problem spaces although these do not represent a new class of problem space. This is shown as a recursion in Figure 1. Finally, we note that a problem space of functions can derive from a problem space of outcomes, as described below.

On the solution space, we distinguish between two classes of solutions. *Operational solutions* are those that can satisfy needs (i.e., solve a problem space of outcomes). *System solutions* are those that can fulfill requirements (i.e., solve a problem of functions). An Operational Solution is the centerpiece by which a problem space of outcomes (Needs) is derived into a problem space of functions (Requirements); the Operational Design consists of a set of open systems and their exchanges, which lead to the desired outcomes, so those exchanging essentially define required functions [5]. As in the case of problem spaces, system solutions can be decomposed and occur at multiple levels of encapsulation, since the class of problem space remains the same. This is shown as a recursive relation in Figure 1. On the other hand, operational solutions are not recursive in the sense of decomposition but can be iteratively developed as the solution is refined. That is, an operational solution can be described in more detail (different abstraction) as operational design decisions are made, but not as parts of an aggregate (decomposition).

We address first the V&V relationships on the problem space. Given that problem spaces are models of problems, we start by conceptualizing traditional model V&V to think about the V&V of Needs and the V&V of requirements. In this sense, we will use the term *verification* to refer to the discovery of knowledge about the element of interest (i.e., being verified) and the term *validation* to refer to the discovery of knowledge about the adequacy and fit-for-purpose of the element of interest (i.e., being validated).

Verification of a problem space addresses therefore the definition of the problem space itself. Because needs and requirements are models, regardless of whether one uses natural language or a different modeling language, verification is assessed against the correct use of the syntactic and semantic rules of the language that is employed. In both cases, the specification of the modeling language that is used to model *Needs* or *Requirements* is extrinsic to the problem and solution layers; note that different modeling languages may be used.

Validation addresses the adequacy of the problem space. For *Needs*, validation addresses the accuracy with which they capture $R_{NEED}$, as a model of the needs is intended to capture what the real needs are. For *Requirements*, validation addresses the extent to which their fulfillment will lead to the satisfaction of *Needs* (for requirements that derive directly from needs) or the fulfillment of requirements at a higher level of encapsulation (for requirements that were decomposed from higher level requirements).

We move now to the V&V of solutions. Following the general conceptualization of verification given earlier, verification of a solution addresses the solution itself: does the solution solve the problem? System verification addresses the fulfillment of *Requirements* (regardless of the level of encapsulation). The verification of an operational solution addresses the fulfillment of *Needs*.

Validation of solutions addresses the satisfaction of the purpose of the problem space, that is, the reason for which the problem space was defined. Looking first a *System solution* for a set of *Requirements* directly derived from a set of *Needs*, the purpose of the *Requirements* is to develop a solution in the space of functions that, when integrated in a context, creates a solution in the space of outcomes that is satisfactory of *Needs*. Therefore, *System validation* addresses the satisfaction of *Needs*. This is different for System solutions that address Requirements that are decomposed from other Requirements (e.g., those traditionally referred to as subsystem, etc.). In this case, the purpose of decomposing the Requirements is to develop smaller solutions than, when integrated, fulfill the Requirements of the larger problem. Therefore, validation of these *System solutions* at lower levels of encapsulation addresses the fulfillment of the *Requirements* on the higher level of encapsulation. (E.g., successful system integration implies subsystem validation.)

Finally, the purpose of *Needs* is to capture $R_{NEED}$. Therefore, *Operational Design validation* addresses the satisfaction of $R_{NEED}$. Furthermore, because the ultimate test of validity is given by the success in the field, we incorporate a second kind of *System validation* against $R_{NEED}$.

## III. Axioms

The conceptual framework has established the key elements and relationships in V&V. To develop this into a formal theory, we first need to establish fundamental axioms about the nature of these activities. These axioms capture essential truths about how verification and validation build knowledge, setting the foundation for our mathematical treatment. To build the underlying theoretical framework, we present three axioms. We believe that they are consistent with reality and do not need further justification for accepting them.

*Axiom 1 (Nature of verification):*
*An agent can collect information about a system.*

*Axiom 2 (Nature of validation):*
*An agent can collect information about the operation of a system with other systems.*

*Axiom 3 (Nature of human interpretation):*
*An agent´s confidence on something is shaped by the information he/she has.*

The theoretical framework is described using Dynamic Epistemic Modal Logic. A summary background is provided in the Appendix.

## IV. PROBLEM AND SOLUTION SPACE ELEMENTS – PRELIMINARIES

Before developing formal definitions of verification and validation, we first specify what exactly is being verified and validated. This section introduces formal definitions of systems, solutions, and related concepts that form the building blocks of our theoretical framework.

*Definition 1 (System solution):*
A System Solution (SysSol) is a set of interrelated elements, where the type of relation is unrestricted, and which transfers energy, matter, or information through its boundaries [adopted from [5]]. This is therefore an open system in General Systems Theory.

*Definition 2 (Operational solution/Context system):*
An Operational Solution (OpsSol) represents the system solution (SysSol) operation in an environment. It encompasses the SysSol, all external systems that directly interact with it (denoted by E), and the interactions between them (denoted by IR). Formally, $OpsSol = (SysSol, E, IR)$. The set formed by SysSol and E is called Context System [5] and is denoted by CS.

*Definition 3 (Outcome):*
An outcome, denoted as o, is a state/consequence/condition of a set of systems that arises from their interaction. The set of all possible outcomes for a set of systems is represented by set $O$, which contains propositions describing these states/consequences/conditions

*Definition 4 (Stakeholder):*
A stakeholder, denoted as S, is a system (called also an actor) that has a vested interest in the SysSol being developed or the project being undertaken or is affected by them. Each stakeholder is associated with a non-empty set of goals, desires, or objectives which attainment is related to the outcomes generated by the CS. Formally, for each stakeholder S, there exists a set of goals $G = \{g_1, g_3, \ldots, g_n\}$, where $n \geq 1$. Each $g_j \in G$ represents a distinct goal, desire, or objective of the stakeholder.

*Definition 5 (Desired Outcomes - $O_d$):*
A set of desired outcomes, denoted by $O_d$, is a set of outcomes that contribute to attaining stakeholder goals. They may involve effects on external systems influenced by the SysSol, whether through direct interaction (i.e., those within the CS) or indirect influence (i.e., those outside of the CS). For each stakeholder goal $g \in G$ of stakeholder S, there exists at least one desired outcome $o_d \in O_d$ such that achieving $o_d$ contributes to the achievement of $g$. It should be noted that these outcomes are not limited to systems within the CS only but also to other systems that are directly influenced by the outcomes of the CS.

## V. A FORMAL LOGIC FOUNDATION FOR V&V THEORY

The choice of formal logic as our mathematical foundation is driven by the fundamental nature of verification and validation activities. At their core, V&V activities involve agents performing specific actions to gain knowledge about system properties and build justified beliefs about requirements fulfillment and needs satisfaction.

Having established the key elements of problem and solution spaces, we need precise mathematical tools that can capture how V&V activities build knowledge about systems. Our conceptual framework identifies the different spaces and relationships involved, but to truly understand verification and validation, we must formalize the subtle ways that evidence leads to beliefs and knowledge.

Dynamic Epistemic Modal Logic (DEML) provides the mathematical machinery needed for this formalization [26, 27]. DEML was specifically developed to model how agents learn and update their knowledge through specific events, which perfectly matches what happens during V&V activities. When an engineer runs a test or performs an analysis, they are executing an event that provides new information, potentially changing their beliefs about the system. Unlike simpler logical frameworks, DEML can capture both the static relationships between different types of knowledge (through modal operators) and the dynamic processes by which knowledge is acquired (through action operators). This combination is essential because verification and validation inherently involve both structural relationships between requirements, needs, and evidence, and processes that generate evidence and build knowledge. Furthermore, DEML's treatment of epistemic uncertainty aligns well with the reality that V&V activities rarely provide absolute certainty but rather build degrees of justified belief about system properties. While other mathematical frameworks like probability theory or classical predicate logic could capture some aspects of V&V, they lack DEML's ability to simultaneously represent knowledge, belief, and the dynamic processes that connect evidence to conclusions.

DEML provides formal tools for representing three fundamental aspects of V&V: how evidence is gathered through specific activities, how that evidence supports beliefs about system properties, and how those beliefs relate to requirements and needs. These align directly with our three axioms about the nature of verification, validation, and human interpretation of evidence introduced in Section III. The complete technical background on DEML is provided in the Appendix.

With this mathematical foundation established, we can develop precise definitions for what makes different V&V artifacts valid and consistent. The formal framework allows us to prove important theorems about the relationships between different V&V activities, the role of evidence in building beliefs about system properties, and the conditions under which verification and validation can overlap or must remain distinct. By providing this theoretical foundation, we resolve longstanding ambiguities in V&V practice and offer systems engineers a structured way to reason about their verification and validation strategies.

## VI. FORMAL BASIS FOR VALIDITY AND CONSISTENCY IN V&V ARTIFACTS

Having established DEML as our mathematical framework, we now develop precise principles for validity and consistency in V&V artifacts. The formal logic foundation allows us to formally define what makes V&V artifacts valid and how they relate to each other through chains of evidence and belief.

### A. Necessity and Sufficiency as Conditions for Validity[1]

From a mathematical standpoint grounded in formal logic, we define the validity of V&V artifacts using *sufficiency* and *necessity*. These two conditions mirror the core logical concepts of implication and contrapositive in DEML, thereby ensuring robust deductive reasoning in our V&V processes.

Sufficiency reflects the principle that a set of premises is adequate to entail a conclusion, formalized as $P \rightarrow Q$ (if P is true, then Q must be true). When applied to V&V artifacts, sufficiency guarantees that when lower-level artifacts are satisfied, they provide enough evidence to establish confidence in the satisfaction of higher-level artifacts. This creates a chain of logical support from concrete evidence through to abstract system properties.

Necessity, on the other hand, captures the contrapositive $\neg Q \rightarrow \neg P$ (if Q is not true, then P is not true), that is if we fail to conclude that a condition is satisfied, then at least one of the necessary premises must be missing or violated. This prevents situations where higher-level properties could be satisfied without proper support from lower-level artifacts, which would indicate gaps in our V&V framework.

Together, sufficiency and necessity establish a strong logical linkage: an "if and only if" connection between artifacts at different levels. This if-and-only-if structure underpins *soundness* (we derive only correct conclusions) and non-vacuousness (we do not claim success without genuinely meeting each artifact). By framing validity in terms of these two conditions, we create a foundation that aligns engineering practice with fundamental logical principles.

---

[1] In this section, validity is not used in the context of V&V in systems engineering, but about the validity of statements in the theory.

### B. Implementation Path

The concepts of necessity and sufficiency must be understood within the context of specific implementation paths, i.e., the different ways we might realize a system to achieve its goals. An implementation path $P$ represents a particular architectural choice, design choice, operational concept, etc., for satisfying stakeholder goals through a specific set of needs, requirements, and V&V artifacts. Understanding implementation paths is crucial because systems may be successfully realized in multiple ways. When we establish sufficiency and necessity relationships between artifacts, these relationships are always relative to a chosen implementation path. What serves as sufficient evidence along one path might be irrelevant or inadequate along another. This path-dependence manifests in how we evaluate belief and knowledge. When we write $B(\Theta) \vDash B(N)$, we are really expressing that the belief in $\Theta$ (e.g., requirements) entails belief in $N$ (e.g., needs) along a specific implementation path. The operator $B_P$ can be used to make this path-dependence explicit, though we omit it in this paper for notational simplicity.

### C. Consistency Implications

While sufficiency and necessity form the core of the validity definition, *consistency* is an implied requirement through DEML's treatment of logical contradictions. In particular, a set of propositions that contains a contradiction ($P \land \neg P$) is impossible to satisfy and cannot coherently satisfy "if and only if" relationships. Likewise, in V&V, an inconsistent set of artifacts would violate any conclusions vacuously (since no conclusion can exist in which all artifacts hold). As a result, no real verification or validation can be claimed in such a contradictory context.

Through formal theorems, we can demonstrate (as seen in the next section) that *consistency* plays a fundamental role in our validity framework, emerging naturally from the interaction between sufficiency and necessity rather than requiring explicit definition. The relationship between consistency and validity manifests in several important ways. First, inconsistency at any level propagates through the artifact chain, corrupting our ability to establish sound logical relationships between different levels of system properties. For example, if requirements contain contradictions, we cannot meaningfully verify whether a system implementation satisfies those requirements, since contradictory requirements cannot be simultaneously satisfied. Second, inconsistency undermines our ability to use V&V evidence effectively –if verification criteria are inconsistent, no amount of testing or verification can establish meaningful conclusions about system behavior. Third, inconsistency in the validation framework can lead to situations where we appear to prove both a property and its negation, destroying our ability to make valid claims about system correctness.

Since consistency emerges naturally from the sufficiency and necessity conditions of validity, it can be claimed that these two conditions (necessity and sufficiency) together, automatically enforce the logical coherence we need for meaningful reasoning about V&V.

### D. Relationships between Validity, Consistency, and Feasibility

In our formal framework, consistency and feasibility are fundamentally equivalent –they represent the same concept viewed through different lenses. Consider two stakeholder desires: "achieve faster-than-light travel" and "complete the solution within 5 years." If we examine these desires purely in terms of formal logic, with no additional context, they appear consistent. One statement speaks of a capability, the other of a timeframe, and there is no immediate logical contradiction between them –unlike saying "the system must be both on and off simultaneously," which is an inherent contradiction. However, if we consider these statements along with physical laws, current technological capabilities, and project execution constraints, we can derive a formal contradiction. This example demonstrates how feasibility and consistency are two sides of the same coin. What makes these stakeholder desires infeasible is precisely what makes them logically inconsistent when we include physical laws in our reasoning. At the surface level there may be no apparent inconsistencies, but when we incorporate physics, technological limitations, and other real-world constraints into our logical framework, we can derive contradictions.

## VII. NEEDS V&V

*Definition 6 (Stakeholder need):*
*A stakeholder need, denoted as η, is a non-recursive logical proposition representing the stakeholder's desired outcomes. It captures what stakeholders expect to achieve through the system's operations (ref. [5]).*

A stakeholder need $\eta$ is a well-formed formula (*wff*) in the logical language $L_d$ (Definition A1 in the Appendix), representing the desired outcomes $O_d \subseteq O$ (Definition 5) associated with stakeholder goals, i.e., the real needs, where: $\eta$ is a logical formula that can be evaluated within the formal system, and $\eta$ adheres to the syntactic and semantic rules of the logical language $L_d$.

*Example:* Without getting any formal semantics, a stakeholder need η can be expressed as follows:
$\eta$ = The <stakeholder S₁> needs <o_d>, where o_d is a desired outcome, such that $o_d \in O_d$

*Definition 7 (Verified Stakeholders' Needs):*
*A set of stakeholder needs, N, is verified if and only if the following condition holds:*
**Well-formedness:** Each need in the set *N* complies with the syntactic and semantic rules of the modeling language used to express the needs. For example, consider the case of using DEML to express needs. A need $n \in N$ is well-formed according to the language $L_d$ (Def. 1) if and only if:
- *n* is an atomic proposition. This condition is to ensure that the needs are singular.
- *n* is a well-formed formula (wff) of $L_d$, i.e., it follows the syntactic rules (see BNF form in Def. 1) of $L_d$.
- *n* is semantically consistent, i.e., there exists a Kripke model $\mathcal{M}$ and a world $w \in W$ such that $\mathcal{M}, w \models n$.

*Definition 8 (Valid Stakeholders' Needs):*
*A set of stakeholder needs, N, along an implementation path, is valid if and only if the following conditions hold:*
  a) *Sufficiency:* The belief in the satisfaction of all needs in N entails the belief in the satisfaction of all stakeholder goals in G. Formally, $B(N) \models B(G)$, where, $B(N) = \bigwedge_{n \in N} B(n)$ represents the combined belief in all needs in N, and $B(G) = \bigwedge_{g \in G} B(g)$ represents the combined belief in all goals in G. Here $\bigwedge$ denotes the logical conjunction.
  b) *Necessity:* The absence of belief in the satisfaction of the needs N entails the absence of belief in the satisfaction of the stakeholder goals G. Formally, $\neg B(N) \models \neg B(G)$. This means that if we do not believe the needs are satisfied, then we cannot believe that the stakeholder goals are satisfied.

*Theorem 1: If a set of stakeholders' needs is inconsistent, then the set of needs cannot be valid.*
<u>Proof:</u> Let N be a set of stakeholder needs that are inconsistent, specifically with conflicting needs $n_1$ and $n_2$. According to Definition 8, for N to be valid, it must satisfy the necessity condition, which states that the absence of belief in the satisfaction of the needs N entails the absence of belief in the satisfaction of the stakeholder goals G. Since N is inconsistent, conflicting needs $n_1$ and $n_2$ cannot be satisfied simultaneously, so one cannot believe that all needs are satisfied making $\neg B(N)$ true. By the necessity condition, this means $\neg B(G)$ is true, so one cannot believe that the stakeholder goals are satisfied. However, the existence of conflicting needs implies that the absence of belief in one need (e.g., $\neg B(n_1)$) allows belief in the conflicting need (e.g., $B(n_2)$), which might satisfy some goals. This contradicts the necessity condition for N because the absence of belief in N does not necessarily prevent belief in the satisfaction of all goals. Therefore, since the necessity condition cannot be satisfied when N is inconsistent, N cannot be valid.

*Theorem 2: If a set of stakeholder needs $\eta$ is inconsistent, then it is impossible to identify system solutions that fully meet all needs in $\eta$ simultaneously.*

*Proof:* Consider a set of stakeholder needs N that is inconsistent, i.e., it contains at least one contradiction. Specifically, there exist needs $\eta_i$ and $\neg\eta_i$ such that both are elements of $N: \eta_i \in N$ and $\neg\eta_i \in N$. This inconsistency leads to a logical contradiction within the set of needs, as the agent believes both $B(\eta_i)$ and $B(\neg\eta_i)$, resulting in $B(\eta_i) \wedge B(\neg\eta_i)$. Now, suppose for the sake of contradiction that there exists a system solution S that fully meets all needs in N. This would imply that S satisfies every need in $N$, including both $\eta_i$ and $\neg\eta_i: S \vDash \eta_i$ and $S \vDash \neg\eta_i$. However, it is logically impossible for any system solution S to simultaneously satisfy a need and its negation. Therefore, the assumption that such a system solution S exists leads to a contradiction. Since no system solution can satisfy both $\eta_i$ and $\neg\eta_i$ simultaneously, it is impossible to identify system solutions that fully meet all needs in the inconsistent set $N$. The inconsistency within $\eta$ prevents any system solution from fulfilling all stakeholder needs at the same time.

*Definition 9 (Minimal Stakeholders' Needs Set):*
A set of stakeholders' needs N is minimal with respect to stakeholder goals G if and only if:
   a) *Validity:* N is valid per Definition 8.
   b) *Essentiality (Non-redundant):* For each need $n \in N$, there exists at least one goal $g \in G$ such that: $B(N) \vDash B(g)$, but $B(N \setminus \{n\}) \nvDash B(g)$. The essentiality condition captures a fundamental property of minimal need sets: every stakeholder need must be truly necessary for satisfying at least one stakeholder goal. When we write $B(N) \vDash B(g), but\ B(N \setminus \{n\}) \nvDash B(g)$, we are expressing that while the complete set of needs N allows us to believe in the satisfaction of goal g, removing need $n$ breaks this ability. This creates a clear test for whether a need is truly essential - if we can remove it without losing our ability to satisfy any stakeholder goal, then it was not really necessary in the first place. This formulation ensures two crucial properties: first, that every need traces to at least one goal (addressing traceability), and second, that each need plays an irreplaceable role in satisfying that goal (ensuring minimality). For example, if removing a need $n$ still allows us to satisfy all goals, then $n$ would be redundant and the set would not be minimal. But if removing $n$ means we can no longer be confident in satisfying even one goal $g$, then $n$ is essential for $g$, justifying its place in our minimal set.

*Theorem 3: Every valid set of stakeholder needs contains at least one minimal set.*
For any valid set of needs N with respect to stakeholder goals G, there exists a minimal set $N_m$ such that $N_m \subseteq N$.
*Proof:* Let N be a valid set of stakeholder needs with respect to goals G. We proceed by considering two exhaustive cases:

   *Case 1 (N is minimal)*: If N is minimal, then by Definition 9, N satisfies both validity and essentiality conditions. Since $N \subseteq N$ (as any set is a subset of itself), N contains a minimal set (itself) as a subset, proving the theorem for this case.

   *Case 2 (N is not minimal)*: If N is not minimal, then while N is valid (by our premise), there must exist at least one need $n \in N$ whose removal preserves validity. Specifically, there exists an $n \in N$ such that both sufficiency $(B(N \setminus \{n\}) \vDash B(G))\ and\ necessity\ (\neg B(N \setminus \{n\}) \vDash \neg B(G))$ still hold when n is removed. Starting with our valid set N, we examine each need $n$ to determine if it can be removed while maintaining validity (Definition 8). Begin with $N_m = N$. For each need $n$ in $N_m$, we test whether $N_m \setminus \{n\}$ remains valid. If removing $n$ preserves both sufficiency $(B(N_m \setminus \{n\}) \vDash B(G))$ and necessity $(\neg B(N_m \setminus \{n\}) \vDash \neg B(G))$, we update $N_m$ by removing $n$. If removing $n$ breaks either condition, we keep $n$ in $N_m$. This process must terminate because we start with a finite set N and can only remove elements a finite number of times. When no more needs can be removed while maintaining validity, the resulting set $N_m$ must be minimal by construction. It satisfies validity because we only remove needs when validity is preserved, and it satisfies essentiality because any attempt to remove a remaining need would break validity for at least one goal.

This theorem implies that a set of stakeholder needs may contain needs that are superfluous and, as such, can be removed from the set without affecting the ability of stakeholders to reach their desired goals.

***Theorem 4: For a set of stakeholder needs, the presence of a minimal set as a subset does not guarantee the validity of the containing set.***
Formally, there exists a set of needs $N_s$ and a minimal set $N_m$ such that $N_m \subseteq N_s$, but $N_s$ is not valid.
<u>Proof:</u> We can prove this by constructing a specific counterexample. Let $N_m$ be a minimal valid set of needs with respect to goals G. We can construct $N_s$ by adding a need $n_x$ to $N_m$ where $n_x$ contradicts some need in $N_m$. Formally: $N_s = N_m \cup \{n_x\}$ where $\exists n \in N_m$ such that $B(n) \vDash \neg B(n_x)$. The set $N_s$ cannot be valid because the contradiction between $n_x$ and n makes it impossible to establish a meaningful sufficiency relationship with the goals. While $B(N_s) \vDash B(G)$ is technically true (as anything follows from a contradiction), this represents a vacuous truth rather than a meaningful entailment. The presence of contradictory needs means we cannot genuinely derive belief in goal satisfaction from our needs set. Thus, we have shown that merely containing a minimal set ($N_m$) is not sufficient for validity, as we can construct a superset ($N_s$) that contains this minimal set but fails to be valid due to the introduction of contradictory needs.

This theorem essentially indicates that any need added to a set of stakeholder needs introduces a risk of making the set invalid, and hence of meeting the goals of the stakeholders.

**VIII. REQUIREMENTS V&V**

*Definition 10 (Requirement):*
*A requirement is a statement θ that defines the required transformation of input to output trajectories by the System Solution (ref. [5, 28]).*
Requirements are only concerned with the SysSol, and not the external systems and their interactions. Requirement statements can be recursive to enable further decomposition.

*Example:* Without getting any formal semantics, a requirement θ can be expressed as follows:
θ = The <SS1> shall <I/O transformation> under <conditions>.

*Definition 11 (Verified Requirements):*
*A set of requirements, Θ, is verified if and only if the following condition holds:*

**Well-formedness:** Each requirement in the set Θ complies with the syntactic and semantic rules of the modeling language used to express the requirements. For example, consider the case of using DEML to express requirements. A requirement θ ∈ Θ is well-formed according to the language $L_d$ (Def. 1) if:
- θ is an atomic proposition. This condition is to ensure that the requirements are singular.
- θ is a well-formed formula (wff) of $L_d$, i.e., it follows the syntactic rules (see BNF form in Def. 1) of $L_d$.
- θ is semantically consistent, i.e., there exists a Kripke model $\mathcal{M}$ and a world $w \in W$ such that $\mathcal{M}, w \vDash \theta$.

*Definition 12 (Valid Requirements):*
*A set of requirements, Θ is s valid if and only if the following conditions hold along an implementation path:*
  a) **Sufficiency:** The belief in the satisfaction of all the requirements in Θ entails the belief in the satisfaction of of all stakeholder needs in N. Formally, $B(\Theta) \vDash B(N)$, where, $B(N) = \bigwedge_{n \in N} B(n)$ represents the combined belief in all needs in N, and $B(\Theta) = \bigwedge_{\theta \in \Theta} B(\theta)$ represents the combined belief in all requirements in Θ. Here ∧ denotes the logical conjunction.
   b) *Necessity:* A set of requirements, Θ, is necessary for a stakeholder need set, η, if and only if the absence of belief in the satisfaction of Θ entails the absence of belief in the satisfaction of η. Formally: $\neg B(\Theta) \vDash \neg B(\eta)$,

*Theorem 5: If a set of requirements is inconsistent, then it cannot be valid.*
*Proof:* This proof is similar to Theorem 1.

*Theorem 6: If a set of stakeholder needs N is inconsistent, then any set of requirements Θ, regardless of its content or consistency, would be considered valid with respect to N*
*Proof:* Let us consider a set of stakeholder needs N that is inconsistent, containing at least one contradiction where both η and ¬η exist in N. This inconsistency leads to $B(\eta)$ and $B(\neg\eta)$, resulting in a contradiction within the belief system. By the principle of explosion in logic, this contradiction allows any proposition to be derived, formally expressed as $B(N) \vdash \bot$ implying $B(N) \vdash B(\varphi)$ for any proposition $\varphi$. The validity of a set of requirements R with respect to N is defined through traceability, sufficiency, necessity, and completeness. Given N's inconsistency, these conditions become trivially satisfied for any R. Traceability is met as N entails any belief, so for any $r \in R$, we can find $\eta \in N$ such that $r$ is traceable to $\eta$. Sufficiency is satisfied as $B(N)$ entails any belief, making $B(R) \vdash B(N)$ trivially true. Necessity is vacuously satisfied due to N's inconsistency. Completeness is trivially met as N's contradictory nature makes it impossible to fail this condition. The inconsistency within N causes all validity conditions to be trivially satisfied by any R, as the logical contradictions in N undermine the meaningful application of these conditions. Therefore, any set of requirements R, regardless of its content or consistency, would be deemed valid with respect to the inconsistent set of stakeholder needs N.

This requirement implies that validation of requirements is insufficient to claim validation of the problem they represent, as the validity of the needs they derive from must be ensured as well.

*Theorem 7: If a set of requirements Θ is inconsistent, then no system solution can satisfy all the requirements in Θ.*
*Proof:* Let us consider a set of requirements R that is inconsistent, meaning it contains at least one contradiction; specifically, there exist requirements $\theta$ and its negation $\neg\theta$ such that both are elements of R. This inconsistency leads to a logical contradiction within the set of requirements, as the system solution would be required to satisfy both $\theta$ and $\neg\theta$ simultaneously. In formal logic, a contradiction implies that there is no possible model in which both $\theta$ and $\neg\theta$ are true at the same time. Therefore, there exists no system solution that can satisfy all the requirements in R, because doing so would necessitate fulfilling mutually exclusive conditions, which is impossible. Consequently, the inconsistency within R precludes the existence of any system solution that satisfies all its requirements, thus proving that if a set of requirements is inconsistent, then no system solution can satisfy all the requirements in R.

This theorem exemplifies the risk of starting and/or maturing a solution against a preliminary set of requirements, as they may constrain the ability of responding to additional requirements derived later. This does not imply though that it is better to wait until all requirements are available, as such assessments depend on the specific conditions of the system's development.

*Definition 13 (Minimal Requirements Set):*
*A set of requirements Θ is minimal with respect to stakeholder needs N if and only if:*
   a) *Validity: Θ is valid per Definition 12.*

b) *Essentiality (Non-Redundant):* For every requirement $\theta \in \Theta$, there exists at least one need $n \in N$ such that: $B(\Theta) \vDash B(n), but\ B(\Theta \setminus \{\theta\}) \nvDash B(n)$. The essentiality condition captures a fundamental property of minimal sets: every requirement must be truly necessary for satisfying at least one stakeholder need.

*Theorem 8: Every valid set of requirements contains at least one minimal set as a subset.*
<u>Proof:</u> This proof is similar to Theorem 3.

*Theorem 9: For a set of requirements, the presence of a minimal set as a subset does not guarantee the validity of the containing set.*
<u>Proof:</u> This proof is similar to Theorem 4.

The same insights to practice stated for Theorems 3 and 4 are applicable here, but in the realm of requirements.

## IX. SYSTEM SOLUTION VERIFICATION

*Definition 14 (Verification criterion):*
*A verification criterion, denoted as $\varphi$, is a statement that defines the measure of success or criteria to judge whether the system solution (SysSol) performs the transformation of input trajectories to output trajectories as specified in the requirement ($\theta$).*

Verification criteria define what needs to be achieved for a requirement to be considered fulfilled. A traditional approach in the field is to define verification methods that must be used (e.g., the mass reported by a test must fulfill the required value in the requirement) [29]. Other approaches demand a set of more precise conditions [30], such as defining the number of points that an optical performance must be measured within a field of view or the precision of such a measurement. This is relevant because there are generally infinite observations that may be done of a system with respect to a requirement (e.g., at different points in time).

*Definition 15 (Valid verification criteria):*
*A set of verification criterion, $\Phi$, along an implementation path, is valid, if and only if the following conditions hold:*
  (a) **Sufficiency:** The belief in the satisfaction of all the verification criteria in $\Phi$ entails the belief in the satisfaction of all requirements in $\Theta$. Formally: $B(\Phi) \vDash B(\Theta)$, where $B(\Theta) = \bigwedge_{\theta \in \Theta} B(\theta)$ represents the combined belief in all requirements in $\Theta$, and $B(\Phi) = \bigwedge_{\varphi \in \Phi} B(\varphi)$ represents the combined belief in all verification criteria in $\Phi$. Here $\bigwedge$ denotes the logical conjunction.
  (b) **Necessity:** The absence of belief in the satisfaction of the verification criteria $\Phi$ entails the absence of belief in the satisfaction of the requirements $\Theta$. Formally: $\neg B(\Phi) \vDash \neg B(\Theta)$

*Theorem 10: If a set of verification criteria $\Phi$ is inconsistent, then the set of verification criteria cannot be valid.*
<u>Proof:</u> This proof is similar in structure to Theorem 1.

*Theorem 11: Addition of new verification criteria to an inconsistent set of verification criteria does not resolve the inconsistency.*
<u>Proof</u>: This proof is similar in structure to Theorem 28.

*Theorem 12: For any consistent set of verification criteria Φ, adding a new criterion or replacing an existing one with a new criterion invalidates any prior guarantee of consistency.*
*Proof:* This proof is similar in structure to Theorem 29.

*Theorem 13: If a set of verification criteria Φ is sufficient to verify a requirement θ, and each criterion in Φ can be decomposed into sub-criteria, then the decomposition of Φ is also sufficient to verify θ.*
*Proof:* Let $\Phi = \{\varphi_1, \varphi_2, \ldots, \varphi_n\}$ be a set of verification criteria sufficient to verify the requirement $\theta$. By the definition of sufficiency of verification criteria, for any epistemic structure M and world w, if the agent believes all the criteria in $\Phi$, then they believe the requirement $\theta$. Formally, this is expressed as $(M, w) \vDash B(\varphi^1 \wedge \varphi^2 \wedge \cdots \wedge \varphi_n) \to (M, w) \vDash B(\theta)$, where $B(\theta)$ denotes the belief that the requirement $\theta$ is satisfied. Assume that each verification criterion $\varphi_i$ in $\Phi$ can be decomposed into a set of sub-criteria, such that $\varphi_i \equiv \varphi_{i1} \wedge \varphi_{i2} \wedge \cdots \wedge \varphi_{im}$, where $\varphi_{ij}$ represents the j-th sub-criterion of the i-th criterion. This equivalence means that believing $\varphi_i$ is the same as believing all its sub-criteria $\varphi_{ij}$, i.e., $(M, w) \vDash B(\varphi_i)$ if and only if $(M, w) \vDash B(\varphi_{i1} \wedge \varphi_{i2} \wedge \cdots \wedge \varphi_{im})$. Since $\Phi$ is sufficient to verify $\theta$, we have $(M, w) \vDash B(\varphi_1 \wedge \varphi_2 \wedge \cdots \wedge \varphi_n) \to (M, w) \vDash B(\theta)$. Substituting each $\varphi_i$ with its decomposition into sub-criteria, we obtain $(M, w) \vDash B((\varphi_{11} \wedge \varphi_{12} \wedge \cdots \wedge \varphi_{1m}) \wedge (\varphi_{21} \wedge \varphi_{22} \wedge \cdots \wedge \varphi_{2m}) \wedge \cdots \wedge (\varphi_{n1} \wedge \varphi_{n2} \wedge \cdots \wedge \varphi_{nm}))$. This expression simplifies to $(M, w) \vDash B(\bigwedge_{i=1}^{n} \bigwedge_{j=1}^{m} \varphi_{ij})$, where $\wedge$ denotes the logical conjunction over all sub-criteria. Combining this result with the sufficiency condition, we have $(M, w) \vDash B(\bigwedge_{i=1}^{n} \bigwedge_{j=1}^{m} \varphi_{ij}) \to (M, w) \vDash B(\theta)$. This means that the agent's belief in all the sub-criteria $\varphi_{ij}$ collectively implies their belief in the requirement $\theta$. Therefore, the decomposition of $\Phi$ into its sub-criteria is also sufficient to verify $\theta$.

*Definition 16 (Minimal Verification Criteria Set):*
A set of verification criteria $\Phi$ is minimal with respect to requirements $\Theta$ if and only if:
   a) *Validity:* $\Phi$ is valid per Definition 15
   b) *Essentiality (Non-Redundant):* For every verification criterion $\varphi \in \Phi$, there exists at least one requirement $\theta \in \Theta$ such that: $B(\Phi) \vDash B(\theta), but\ B(\Phi \setminus \{\varphi\}) \nvDash B(\theta)$.

*Theorem 14: Every valid set of verification criteria contains at least one minimal set as a subset.*
*Proof:* This proof is similar to Theorem 3.

*Theorem 15: For a set of verification criteria, the presence of a minimal set as a subset does not guarantee the validity of the containing set.*
*Proof:* This proof is similar to Theorem 4.

The same insights as for Theorems 3 and 4 apply here.

*Theorem 16: If a verification criterion φ in a set of verification criteria Φ is not essential (as defined in the minimal set definition), then its removal does not affect the verification of the System Solution with respect to the requirements.*
*Proof:* Let $\Phi$ be a set of verification criteria sufficient to verify the requirements $\Theta$. Let $\varphi$ be a non-essential criterion in $\Phi$, meaning that for every requirement $\theta \in \Theta, if\ B(\Phi) \vDash B(\theta), then\ also\ B(\Phi \setminus \{\varphi\}) \vDash B(\theta)$. This is precisely the negation of the essentiality condition from our minimal set definition, which states that an essential criterion must break verification of at least one requirement when removed. Our goal is to show that removing $\varphi$ does not affect the verification of the System Solution with respect to $\Theta$. By the sufficiency condition of validity, we have $B(\Phi) \vDash B(\Theta)$. Since $\varphi$ is non-essential, for each $\theta \in \Theta$, we have $B(\Phi \setminus \{\varphi\}) \vDash B(\theta)$. Taking the conjunction over all $\theta \in \Theta$, this means $B(\Phi \setminus \{\varphi\}) \vDash B(\Theta)$. By the necessity condition of validity, we also have $\neg B(\Phi) \vDash \neg B(\Theta)$. Since $\varphi$ is non-essential, the contrapositive of our earlier result shows that $\neg B(\Phi \setminus \{\varphi\}) \models \neg B(\Theta)$. Therefore, $\Phi \setminus \{\varphi\}$ maintains both sufficiency and necessity with respect to $\Theta$, which means the removal of $\varphi$ does not affect the verification of the System Solution with respect to the requirements.

### Definition 17 (Verification activity):
An event $v \in V$ is a verification activity after which an agent acquires knowledge about the verification evidence, $\varphi_e$ (Definition 18). Here, V is the set of all verification activities. A verification activity is represented in the dynamic epistemic language using the dynamic modal operator (as defined in Eqs. A7-A9 in the Appendix), specifically, $[v]$ followed by the verification criteria. For example, $[v]K_A(\varphi_e \vee \neg\varphi_e)$ means that "after event v takes place, agent A knows whether $\varphi_e$ holds."
Mathematically, "$(M, w)$ satisfies the propositions that after the verification activity $v$ takes place, $\varphi_e$ holds" is defined as follows: $(M, w) \models [v](\varphi_e) \Leftrightarrow (M, w) \models [v] \rightarrow \varphi_e$

### Definition 18 (Verification evidence):
Verification evidence, denoted as $\varphi_e$, is a statement that represents the data or information obtained as a direct result of performing a verification activity $v$ (Definition 17).

The acquisition of the verification evidence, i.e., the knowledge about the verification evidence $\varphi_e$ leads to the belief about the truth or falsity of the associated verification criterion $\varphi$.
To state this formally: After performing the verification activity v, the agent knows the truth or falsity of the verification evidence, i.e., $[v]K(\varphi_e)$ or $[v]K(\neg\varphi_e)$, where $K(\varphi_e) \rightarrow B(\varphi)$ and $K(\neg\varphi_e) \rightarrow B(\neg\varphi)$.

      This definition formalizes the concept of verification evidence as information resulting from verification activities. The notation $\phi_e$ represents this evidence, linking it directly to the verification criterion $\phi$. By defining verification evidence in this way, we establish a clear connection between the activities performed ($v$) and the knowledge gained ($K(\phi_e)$). This approach allows us to model how the performance of verification activities leads to knowledge about the system, which in turn informs beliefs about the verification criteria. The implication $K(\phi_e) \rightarrow B(\phi)$ captures the crucial step where tangible evidence (such as looking at the results of a test) translates into beliefs about the verification criteria (whether those results, and the underlying test, are sufficient to consider the requirement fulfilled).

      The relationship between verification evidence and the agent's assessment is critical. While the evidence itself is tangible data obtained from verification activities, its interpretation and the resulting beliefs about the verification criteria are specific to the assessing agent. This connection between evidence and agent assessment is formalized in Definition 21 (Verification assessment), where we elaborate on how agents form beliefs about verification criteria based on observed evidence.

### Definition 19 (Valid verification activity):
A set of verification activities V along an implementation path is valid if and only if the following conditions hold:
  (a) **Sufficiency:** The combined knowledge gained from performing all verification activities in V about their respective verification evidence $\varphi_e$ entails the belief in the satisfaction of all verification criteria in $\Phi$. Formally: $K_V(\Phi_e) \models B(\Phi)$ where: $K_V(\Phi_e) = \wedge_{v \in V}[v]K(\varphi_e)$, $\Phi_e$ is the set of all verification evidence obtained from V, and $B(\Phi) = \wedge_{\varphi \in \Phi} B(\varphi)$. Here $\wedge$ denotes the logical conjunction.
  (b) **Necessity:** The absence of the combined knowledge from the verification activities in V about the verification evidence $\Phi_e$ entails the absence of belief in the satisfaction of the verification criteria $\Phi$. Formally: $\neg K_V(\Phi_e) \models \neg B(\Phi)$.

*Theorem 17: If a set of verification activities is inconsistent, then the set of verification activities cannot be valid.*
*Proof:* This proof is similar in structure to Theorem 1.

Inconsistency here does not refer to inconsistency of the evidence stemming from the verification activities, which could be resolved by conducting additional activities to collect additional evidence, but about inconsistencies in the activities themselves. For example, one would be to request a test on a system that requires the full integrity of the system after it has gone through a successfully executed destructive test.

*Theorem 18: Addition of new verification activities to an inconsistent set of verification activities does not resolve the inconsistency.*
*Proof:* Let $V$ be an inconsistent set of verification activities. Assume, for the sake of contradiction, that adding a new verification activity $v$ to $V$ resolves the inconsistency. Let $V' = V \cup \{v\}$ be the new set of verification activities. For the set of verification activities to be consistent (Definition A7 in the Appendix), V' must be conjointly satisfiable, i.e., there exists an epistemic structure M and a world w such that $(M, w) \vDash V'$. Since $V' = V \cup \{v\}$, and $(M, w) \vDash V'$, it follows that $(M, w) \vDash V$. However, this contradicts the initial premise that V is inconsistent. By the definition of consistency (Definition A7 in the Appendix ), if V is inconsistent, there cannot exist any epistemic structure M and world w such that $(M, w) \vDash V$. This contradiction shows that the earlier assumption that the addition of a new verification activity $v$ to $V$ resolves the inconsistency must be false.

*Theorem 19: If two sets of valid verification activities $V_1$ and $V_2$ generate equivalent knowledge about the verification evidence for all verification criteria in $\Phi$, then these sets can be used interchangeably without affecting the verification of $\Phi$.*
*Proof:* Let $V_1$ and $V_2$ be two sets of valid verification activities, and $\Phi$ be the set of verification criteria. Assume $V_1$ and $V_2$ generate equivalent knowledge about $\Phi_e$ (verification evidence), i.e., for any epistemic structure M and world w, $(M, w) \vDash K_{V_1}(\Phi_e)$ iff $(M, w) \vDash K_{V_2}(\Phi_e)$. By the sufficiency condition in Definition 19, we have $K_{V_1}(\Phi_e) \vDash B(\Phi)$ and $K_{V_2}(\Phi_e) \vDash B(\Phi)$. Given the equivalence of knowledge generated by $V_1$ and $V_2$, we can conclude that they lead to the same beliefs about $\Phi$. The same argument holds for $\neg \Phi_e$ and $\neg B(\Phi)$. Therefore, $V_1$ and $V_2$ can be substituted for each other without affecting the verification of $\Phi$.

*Theorem 20: If a set of verification activities V is valid, then the beliefs formed about the verification criteria $\Phi$ are necessarily derived from the knowledge generated by the verification evidence produced by V.*
*Proof:* Let V be a valid set of verification activities and $\Phi$ be the set of verification criteria. By the sufficiency condition in Definition 19, we have $K_V(\Phi_e) \vDash B(\Phi)$. This directly states that the knowledge generated by V about the verification evidence $\Phi_e$ leads to the belief in $\Phi$. Moreover, by the necessity condition, we have $\neg K_V(\Phi_e) \vDash \neg B(\Phi)$. This means that the absence of knowledge about the verification evidence necessarily leads to the absence of belief in the verification criteria. Together, these conditions demonstrate that the beliefs about the verification criteria $\Phi$ are necessarily derived from the knowledge generated by V. The sufficiency condition ensures that the knowledge is adequate to form the beliefs, while the necessity condition ensures that these beliefs cannot be formed without the knowledge generated by V. Therefore, we can conclude that if V is a valid set of verification activities, the beliefs formed about $\Phi$ are necessarily derived from the knowledge generated by the verification evidence produced by V.

*Definition 20 (Minimal Verification Activity Set):*
*A set of verification activities V is minimal with respect to verification evidence $\Phi_e$ if and only if:*

a) *Validity:* V is valid per Definition 19 (Valid Verification Activity)
b) *Essentiality (Non-Redundant):* For every verification activity $v \in V$, there exists a verification criterion $\varphi \in \Phi$ such that:
   - The execution of $v$ leads to knowledge about verification evidence $\varphi_e$
   - The belief in this evidence entails belief in the criterion
   - Removing $v$ breaks this relationship, i.e., $\forall v \in V, \exists \varphi \in \Phi$ such that: $([v]K(\varphi_e) \wedge B(\varphi_e) \vDash B(\varphi))$ but $\neg \exists v_2 \in (V \setminus \{v\})$ such that $[v_2]K(\varphi_e) \wedge B(\varphi_e) \vDash B(\varphi)$

*Theorem 21: Every valid set of verification activities contains at least one minimal set as a subset.*
Formally, for any valid set of verification activities $V$ with respect to verification evidence $\Phi_e$ and criteria $\Phi$, there exists a minimal set $V_m$ such that $V_m \subseteq V$, where for each $v \in V_m$, there exists $\varphi \in \Phi$ such that $([v]K(\varphi_e) \wedge B(\varphi_e) \vDash B(\varphi))$ but $\neg \exists v_x \in (V_m \setminus \{v\})$ such that $[v_x]K(\varphi_e) \wedge B(\varphi_e) \vDash B(\varphi)$.
*Proof:* Let V be a valid set of verification activities. We can construct a minimal set $V_m$ through an iterative reduction process. Starting with $V_m = V$, we examine each activity $v$ in $V_m$ to determine if other activities in the set can provide the same evidence chain to criteria. For each activity $v$, we check if there exists any other activity $v_x$ in $V_m \setminus \{v\}$ that can establish the same evidence-to-criterion relationships, i.e., where $[v_x]K(\varphi_e) \wedge B(\varphi_e) \vDash B(\varphi)$ for all $\varphi$ that $v$ supports. If such a $v_x$ exists, we can remove $v$ while maintaining all necessary evidence chains. If no such $v_x$ exists, $v$ must be kept as it provides unique essential evidence for at least one criterion. This process must terminate since we start with a finite set $V$ and can only remove elements a finite number of times. When no more activities can be removed while maintaining valid evidence chains, the resulting set $V_m$ must be minimal by construction.

*Theorem 22: For a set of verification activities, the presence of a minimal set as a subset does not guarantee the validity of the containing set.*
Formally, there exists a set of verification activities $V_s$ and a minimal set $V_m$ such that $V_m \subseteq V_s$, but $V_s$ is not valid.
*Proof:* We can prove this by constructing a specific counterexample. Let $V_m$ be a minimal valid set of verification activities. We can construct $V_s$ by adding an activity $v'$ to $V_m$ where $v'$ leads to knowledge $[v']K(\varphi_e')$ such that $B(\varphi_e') \vDash \neg B(\varphi)$ for some $\varphi \in \Phi$ that is supported by activities in $V_m$. Formally, $V_s = V_m \cup \{v'\}$ where $v'$ provides evidence that contradicts the evidence chain for some criterion in $\Phi$. The resulting set $V_s$ cannot be valid because it contains activities that lead to contradictory evidence about the same criterion, despite containing the minimal set $V_m$

*Theorem 23: If a verification activity $v$ in $V$ does not belong to the minimal set, then its removal does not affect the verification of any criteria.*
*Proof:* Let $v$ be a non-essential activity in $V$. By our definition of essentiality, for each criterion $\varphi$ that $v$ provides evidence for, there exists some other activity $v_x \in V \setminus \{v\}$ such that $[v_x]K(\varphi_e) \wedge B(\varphi_e) \vDash B(\varphi)$. This means that for every criterion $\varphi$ that $v$ contributes to verifying, there exists an alternative activity that maintains the necessary evidence chain. Therefore, removing $v$ preserves both the sufficiency relationship $K(V \setminus \{v\}) \vDash K(\Phi_e)$ and the necessity relationship $\neg K(V \setminus \{v\}) \vDash \neg K(\Phi_e)$. Since no unique evidence-to-criterion relationships are broken by removing $v$, the verification of all criteria remains intact. Thus, removing a non-essential activity does not affect the overall verification of the system with respect to the criteria.

*Definition 21 (Verification assessment):*
*A verification Assessment, denoted as $A_{\varphi_e}^{ver}$, is the agent's belief about the value of the verification criterion $\varphi$ based on the observed state of the verification evidence $\varphi_e$, which can take on multiple values $\{\varphi_{e1}, \varphi_{e2}, \ldots, \varphi_{en}\}$.*

Formally, $A_{\varphi_e}^{ver} := \bigwedge_{i=1}^{n} B(\varphi_{ei} \rightarrow (\varphi = x_i))$, where, B is the belief operator; $\varphi_e$ is the verification evidence; $\varphi_{ei}$ is the *i-th* state of $\varphi_e$; $\varphi = x_i$ represents the agent assigning a specific value $x_i$ to the verification criterion $\varphi$ when $\varphi_e$ is observed to be in state $\varphi_{ei}$; $x_i$ are specific values (e.g., numerical values, qualitative assessments) corresponding to each state $\varphi_{ei}$. As an example, the verification assessment $A_{\varphi_{e1}}^{ver}$ is the agent's belief that if the verification evidence $\varphi_{e1}$ holds, then the verification criterion $\varphi$ takes the value $x_1$.

The verification assessment, $A_{\varphi_e}^{ver}$, is defined using a conjunction of beliefs about implications from evidence states to criterion values. This formulation allows us to represent the agent's complete assessment of how different evidence states relate to verification outcomes. By using implications within belief operators $(B(\phi_{ei} \rightarrow (\phi = x_i)))$, we can model conditional relationships without needing explicit conditional probability operators. This approach provides a flexible framework for representing complex verification scenarios, where different states of evidence may lead to different assessments of the verification criteria.

***Theorem 24: If two verification criteria are logically independent, then the order in which they are verified does not affect the overall verification assessment.***

*Proof:* Let $\varphi_1$ and $\varphi_2$ be two logically independent verification criteria, meaning $B(\varphi_1) \not\vDash B(\varphi_2)$, $B(\varphi_2) \not\vDash B(\varphi_1)$, $B(\varphi_1) \not\vDash B(\neg\varphi_2)$, and $B(\varphi_2) \not\vDash B(\neg\varphi_1)$, where B represents the belief operator and $\not\vDash$ denotes "does not entail." Let $v_1$ and $v_2$ be the corresponding verification activities, yielding evidence $\varphi_{e1}$ and $\varphi_{e2}$. We consider two verification sequences: (1) $v_1$ followed by $v_2$, and (2) $v_2$ followed by $v_1$. In sequence 1, the agent first acquires knowledge $[v_1]K(\varphi_{e1})$, forming belief $B(\varphi_1)$, then $[v_2]K(\varphi_{e2})$, forming $B(\varphi_2)$. The overall assessment is $B(\Phi) = B(\varphi_1) \wedge B(\varphi_2)$. In sequence 2, the order is reversed, resulting in $B(\Phi') = B(\varphi_2) \wedge B(\varphi_1)$. Due to the logical independence of $\varphi_1$ and $\varphi_2$, the belief formed about one criterion does not influence the other. Moreover, the commutativity of logical conjunction ensures that $B(\Phi) = B(\varphi_1) \wedge B(\varphi_2) = B(\varphi_2) \wedge B(\varphi_1) = B(\Phi')$. Thus, the order of verification does not affect the final belief state or the overall verification assessment, as the agent's combined belief in the criteria remains the same regardless of the sequence of verification activities.

*Definition 22 (Verification confidence):*

*Verification Confidence, denoted as $C_{A_{\varphi_e}^{ver}}$, is the degree of belief held by an agent about the verification assessment $A_{\varphi_e}^{ver}$. It represents the agent's confidence in their assessment of the verification criterion $\varphi$ based on the observed states of verification evidence $\varphi_e$.*

Formally, for each possible state $\varphi_{ei}$ of the verification evidence $\varphi_e$ and the corresponding assessment value $x_i$: $C_{A_{\varphi_e}^{ver}} := B((\varphi_{ei} \rightarrow (\varphi = x_i)) \rightarrow c_{xi}) \wedge B(\neg(\varphi_{ei} \rightarrow (\varphi = x_i)) \rightarrow c_{yi})$, where: $c_{xi}$ represents the proposition "The confidence in the assessment is $x_i$;

$c_{yi}$ represents the proposition "The confidence in the negation of the assessment is $y_i$";

$x_i$ and $y_i$ are numbers between 0 and 1, with $x_i + y_i = 1$. As an example, the verification confidence $C_{A_{\varphi_{e1}}^{ver}}$ is the agent's belief that if the verification evidence $\varphi_{e1}$ holds, then the confidence in the assessment is $c_{x1}$, and if the verification evidence $\varphi_{e1}$ does not hold, then the confidence in the negation of the assessment is $c_{y1}$.

This is defined to capture the degree of belief in verification assessments, representing the agent's certainty about their judgments. By defining confidence for both the assessment and its negation ($c_{xi}$ and $c_{yi}$), we account for the full spectrum of belief, from certainty to complete doubt. The constraint $x_i + y_i = 1$ ensures that confidence levels are complementary, reflecting the idea that increased confidence in an assessment necessarily decreases confidence in its negation. This definition allows for nuanced representation of confidence in verification outcomes, which is crucial in real-world scenarios where absolute certainty is rare. By formalizing confidence in this way, we enable reasoning about the reliability of verification results, which is essential for risk assessment and decision-making in system development and deployment.

### Definition 23 (Valid verification confidence):
A verification confidence, $C_{A_{\varphi_e}^{ver}}$, is considered valid if it satisfies the following conditions for each possible state $\varphi_{ei}$ of the verification evidence $\varphi_e$:
- The agent knows the observed state of the verification evidence: $K(\varphi_{ei})$
- **Uniqueness and Completeness:** For each verification assessment corresponding to $\varphi_{ei}$, there exists exactly one pair of values $(x_i, y_i)$ such that $x_i + y_i = 1$.
- **Certainty:** The verification confidence is 1 (i.e., $x_i = 1, y_i = 0$) if and only if: $B(\varphi_{ei} \Leftrightarrow (\varphi = x_i)) \wedge K(\varphi_{ei})$
- **Impossibility:** The verification confidence is 0 (i.e., $x_i = 0, y_i = 1$) if and only if: $B(\varphi_{ei} \to (\varphi \neq x_i)) \wedge K(\varphi_{ei})$
- **Range:** The verification confidence is between 0 and 1 (i.e., $0 < x_i < 1, 0 < y_i < 1$) if and only if: $B(\varphi_{ei} \to (\varphi = x_i)) \wedge K(\varphi_{ei}) \wedge \neg B(\varphi_{ei} \Leftrightarrow (\varphi = x_i))$
- **Non-Contradictory:** The agent cannot simultaneously believe that $\varphi_{ei}$ implies $\varphi = x_i$ and that $\varphi_{ei}$ implies $\varphi \neq x_i$, i.e., $\neg(B(\varphi_{ei} \to (\varphi = x_i)) \wedge B(\varphi_{ei} \to (\varphi \neq x_i)))$

### Definition 24 (Verified System solution):
A system solution, SysSol, is considered verified if and only if,
a) For all requirements $\theta \in \Theta$, there exists a set of verification criteria $\Phi$ where:
   - $B(\Phi) \vDash B(\theta)$ holds (sufficiency)
   - $\neg B(\Phi) \vDash \neg B(\theta)$ holds (necessity)
b) Let $A^{ver}$ be the set of verification assessments and $C_{A_{\varphi_{ei}}^{ver}}$ be the set of verification confidences. For each verification criterion $\varphi \in \Phi$ and its associated verification evidence $\varphi_e$, the system is verified if and only if: $\forall A_{\varphi_{ei}}^{ver} \in A^{ver}: (C_{A_{\varphi_{ei}}^{ver}} \geq \tau)$
c) Knowledge of verification evidence implies belief in the verification criterion: $K(\varphi_e) \to B(\varphi)$

### Theorem 25: Adding new verification evidence to an existing set of evidence cannot decrease the confidence in any verification criterion, as long as the new evidence is consistent with the existing evidence.

<u>Proof</u>: Let $A_{\varphi_e}^{ver} := \bigwedge_{i=1}^{n} B(\varphi_{ei} \to (\varphi = x_i))$ be the verification assessment based on the existing set of verification evidence $\varphi_e = \{\varphi_{e1}, \varphi_{e2}, \ldots, \varphi_{en}\}$, where $\varphi_{ei}$ are the states of the verification evidence, and $x_i$ are the corresponding values assigned to the verification criterion $\varphi$. Let $C_{A_{\varphi_e}^{ver}}$ be the corresponding verification confidence, defined as: $C_{A_{\varphi_e}^{ver}} := \bigwedge_{i=1}^{n}[B((\varphi_{ei} \to (\varphi = x_i)) \to c_{xi}) \wedge B(\neg(\varphi_{ei} \to (\varphi = x_i)) \to c_{yi})]$ where $c_{xi}$ and $c_{yi}$ are confidence levels satisfying $x_i + y_i = 1$. Now, suppose we have new evidence $e$ that is consistent with the existing evidence $\varphi_e$ and does not reduce the relevance to the verification criterion $\varphi$. Let $\varphi_e' = \varphi_e \cup \{e\}$ be the updated set of evidence. We aim to show that the new verification confidence $C_{A_{\varphi_e}^{ver}}'$ satisfies $c_{xi}' \geq c_{xi}$ for all $i$, that is, that the confidence in any verification criterion does not decrease. We consider three cases based on the initial confidence levels:

    *Case 1 - Certainty ($c_{xi}$ is 1):* This implies that $B(\varphi_{ei} \Leftrightarrow (\varphi = x_i)) \wedge K(\varphi_{ei})$. Since the new evidence $e$ is consistent with $\varphi_{ei}$ and does not reduce relevance, the agent continues to believe the equivalence and knows $\varphi_{ei} \cup \{e\}$: $B((\varphi_{ei} \wedge e) \Leftrightarrow (\varphi = x_i)) \wedge K(\varphi_{ei} \cup \{e\})$. By the certainty condition in Definition 30, the updated confidence remains $c_{xi}' = 1$, so $c_{xi}' = c_{xi}$.

    *Case 2 - Impossibility ($c_{xi}$ is 0):* This implies that: $B(\varphi_{ei} \to (\varphi \neq x_i)) \wedge K(\varphi_{ei})$. Since the new evidence is consistent with $\varphi_{ei}$, this belief remains unchanged: $B((\varphi_{ei} \wedge e) \to (\varphi \neq x_i)) \wedge K(\varphi_{ei} \cup \{e\})$. So, by the impossibility condition in Definition 30, the updated confidence remains $c_{xi}' = 0$, so $c_{xi}' = c_{xi}$.

Case 3 - Partial Confidence (0 < $c_{xi}$ < 1): Here the agent believes: $B(\varphi_{ei} \rightarrow (\varphi = x_i)) \wedge K(\varphi_{ei}) \wedge \neg B(\varphi_{ei} \Leftrightarrow (\varphi = x_i))$. Adding consistent evidence $e$ can increase the relevance if $e$ provides additional support, where the agent's belief in $\varphi = x_i$ strengthens, leading to $c_{xi}' > c_{xi}$. On the other hand, it can maintain relevance if $e$ does not affect relevance, leading to the confidence being the same, $c_{xi}' = c_{xi}$. Since $e$ is consistent and does not reduce relevance, the updated confidence satisfies $c_{xi}' \geq c_{xi}$.

In all cases, adding new, consistent evidence $e$ cannot decrease the confidence in any verification criterion. Therefore, $c_{xi}' \geq c_{xi}$ for all $i$.

## X. SYSTEM SOLUTION VALIDATION

*Definition 25 (Validation criterion):*
*A validation criterion, denoted as $\psi$, is a statement that defines the measure of success or criteria to judge whether the interactions between the system solution (SysSol) and the external systems within the context system produce the desired outcomes that satisfy the stakeholder needs ($\eta$).*

*Definition 26 (Valid validation criteria):*
*A set of validation criteria, $\psi$, is valid, if and only if the following conditions hold:*
  a) ***Sufficiency:*** The belief in the satisfaction of all the validation criteria in $\Psi$ entails the belief in the satisfaction of all stakeholder needs in N. Formally, $B(\Psi) \vDash B(N)$, where $B(N) = \bigwedge_{\eta \in N} B(\eta)$ represents the combined belief in all needs in N, and $B(\Psi) = \bigwedge_{\psi \in \Psi} B(\psi)$ represents the combined belief in all validation criteria in $\Psi$.
  b) ***Necessity:*** The absence of belief in the satisfaction of the validation criteria $\Psi$ entails the absence of belief in the satisfaction of the stakeholder needs N. Formally: $\neg B(\Psi) \vDash \neg B(N)$

***Theorem 26: If a set of validation criteria $\Psi$ is inconsistent, then the set of validation criteria cannot be valid.***
*Proof:* This proof is similar in structure to Theorem 1.

***Theorem 27: Addition of new validation criteria to an inconsistent set of validation criteria does not resolve the inconsistency.***
*Proof*: This is trivial. An inconsistent set of validation criteria means $(M, w) \vDash \Psi$ is empty, where $\Psi$ is the conjunction of all validation criteria. Even addition of a new validation criterion ($\psi_n$) that is tautologous still maintains $(M, w) \vDash \Psi \wedge \psi_n$ being empty, given that the conjunction operation is used.

***Theorem 28: For any consistent set of validation criteria $\Psi$, adding a new criterion or replacing an existing one with a new criterion invalidates any prior guarantee of consistency.***
*Proof:* Let $\Psi = \{\psi_1, \psi_2, \ldots, \psi_n\}$ be a consistent set of validation criteria. Let $\varphi$ be a new criterion we wish to add or use as a replacement. Consider two possible scenarios: (1) $\varphi$ is logically independent of all $\psi_i$ in $\Psi$. In this case, $\Psi \cup \{\varphi\}$ might be consistent, but we cannot know without checking. (2) $\varphi$ logically contradicts some subset of $\Psi$. For example, if $\varphi = p$ and there exists a $\psi_i = \neg p$ in $\Psi$, then $\Psi \cup \{\varphi\}$ is inconsistent. Similarly, replacing $\psi_i$ with $\varphi$ in this case would also result in an inconsistent set. Since we cannot determine a priori which scenario applies to $\varphi$, a new consistency check is necessary for both addition and replacement operations to ensure the resulting set of validation criteria remains consistent.

**Theorem 29**: *If a set of validation criteria $\Psi$ is sufficient to validate a stakeholder need $\eta$, and each criterion in $\Psi$ can be decomposed into sub-criteria, then the decomposition of $\Psi$ is also sufficient to validate $\eta$.*

*Proof*: Let $\Psi = \{\psi_1, \psi_2, \ldots, \psi_n\}$ be a set of validation criteria sufficient to validate whether a system solution meets the stakeholder need $\eta$. By the definition of sufficiency of verification criteria (Definition 15), for any epistemic structure M and world w, if $(M, w) \vDash B(\psi_1 \wedge \psi_2 \wedge \ldots \wedge \psi_n)$, then $(M, w) \vDash B(\eta)$, where $B(\eta)$ represents the belief that the system solution meets the stakeholder need $\eta$. Now, consider that each criterion $\psi_i$ can be decomposed into a set of sub-criteria. We can express this as an equivalence: $\psi_i \equiv \psi_{i1} \wedge \psi_{i2} \wedge \ldots \wedge \psi_{im}$, where $\psi_{ij}$ represents the j-th sub-criterion of the i-th criterion. This equivalence means that believing a criterion is satisfied is the same as believing all of its sub-criteria are satisfied: $(M, w) \vDash B(\psi_i)$ if and only if $(M, w) \vDash B(\psi_{i1} \wedge \psi_{i2} \wedge \ldots \wedge \psi_{im})$. We can apply this decomposition to all criteria in $\Psi$. After substitution, the belief in $\Psi$ becomes equivalent to the belief in all sub-criteria: $(M, w) \vDash B((\psi_{11} \wedge \psi_{12} \wedge \ldots \wedge \psi_{1m}) \wedge (\psi_{21} \wedge \ldots \wedge \psi_{nm}))$. This can be simplified to $(M, w) \vDash B(\psi_{11} \wedge \psi_{12} \wedge \ldots \wedge \psi_{nm})$. Since we know that belief in $\Psi$ implies belief in $\eta$ (i.e., belief that the system solution meets the stakeholder need), and we've shown that belief in all sub-criteria is equivalent to belief in $\Psi$, we can conclude that belief in all sub-criteria also implies belief that the system solution meets $\eta$. Therefore, the set of all sub-criteria, $\Psi' = \{\psi_{11}, \psi_{12}, \ldots, \psi_{nm}\}$, is sufficient to validate whether the system solution meets the stakeholder need $\eta$.

*Definition 27 (Minimal Validation Criteria Set):*
*A set of validation criteria $\Psi$ is minimal with respect to stakeholder needs N if and only if:*
  a) **Validity:** $\Psi$ is valid per Definition (Valid Validation Criteria)
  b) **Essentiality (Non-Redundant):** For every validation criterion $\psi \in \Psi$, there exists at least one need $\eta \in N$ such that: $B(\Psi) \vDash B(\eta), \text{but } B(\Psi \setminus \{\psi\}) \nvDash B(\eta)$. The essentiality condition captures a fundamental property of minimal sets: every validation criterion must be truly necessary for validating at least one stakeholder need. When we write $B(\Psi) \vDash B(\eta), \text{but } B(\Psi \setminus \{\psi\}) \nvDash B(\eta)$, we are expressing that while the complete set of validation criteria $\Psi$ allows us to believe in the satisfaction of need $\eta$, removing criterion $\psi$ breaks this ability. This creates a clear test for whether a criterion is truly essential - if we can remove it without losing our ability to validate any stakeholder need, then it was not really necessary in the first place. This formulation ensures both that every validation criterion traces to at least one need and that each criterion plays an irreplaceable role in validating that need.

**Theorem 30**: *Every valid set of validation criteria contains at least one minimal set as a subset*
Proof: This proof is similar to Theorem 3.

**Theorem 31**: *For a set of validation criteria, the presence of a minimal set as a subset does not guarantee the validity of the containing set*
Proof: This proof is similar to Theorem 4.

**Theorem 32**: *If a validation criterion $\psi$ in a set of verification criteria $\Psi$ is not essential (as defined in the minimal set definition), then its removal does not affect the validation of the System Solution with respect to the stakeholders' needs.*
Proof: This proof is similar to Theorem 16.

*Definition 28 (Validation activity):*
*An event $u \in U$ is a validation activity after which an agent acquires knowledge about the validation evidence, $\psi_e$ (Definition 29). Here U is the set of all validation activities. A validation activity is represented in the dynamic epistemic language using the dynamic modal operator, specifically, $[u]$ followed by the validation criteria. For example, $[u]K(\psi_e \vee \neg\psi_e)$ means that "after event u takes place, the agent knows whether $\psi_e$ holds".*

*Definition 29 (Validation evidence):*
*Validation evidence, denoted as $\psi_e$, is a proposition that represents the objective data or information obtained as a direct result of performing a validation activity u (Definition 28).*

*The acquisition of the validation evidence, i.e., the knowledge about the validation evidence $\psi_e$ leads the agent to form beliefs about the truth or falsity of the associated validation criterion $\psi$.*
To state this formally: After performing the validation activity, u, the agent knows the truth or falsity of the validation evidence, i.e., $[u]K(\psi_e)$ or $[u]K(\neg\psi_e)$, where $K(\psi_e) \to B(\psi)$, and $K(\neg\psi_e) \to B(\neg\psi)$

*Definition 30 (Valid validation activity):*
*A set of validation activities U along an implementation path is valid if and only if the following conditions hold:*
  a) **Sufficiency:** The combined knowledge gained from performing all validation activities in U about their respective validation evidence $\psi_e$ entails the belief in the satisfaction of all validation criteria in $\Psi$. Formally: $K_U(\Psi_e) \vDash B(\Psi)$ where: $K_U(\Psi_e) = \bigwedge_{u \in U}[u]K(\psi_e)$, $\Psi_e$ is the set of all validation evidence obtained from U, and $B(\Psi) = \bigwedge_{\psi \in \Psi} B(\psi)$
  b) **Necessity:** The absence of the combined knowledge from the validation activities in U about the validation evidence $\Psi_e$ entails the absence of belief in the satisfaction of the validation criteria $\Psi$. Formally: $\neg K_U(\Psi_e) \vDash \neg B(\Psi)$.

**Theorem 33: *If a set of validation activities is inconsistent, then the set of validation activities cannot be valid.***
<u>Proof:</u> This proof is similar in structure to Theorem 1.

**Theorem 34: *Addition of new validation activities to an inconsistent set of validation activities does not resolve the inconsistency.***
<u>Proof:</u> Let $U$ be an inconsistent set of validation activities. Assume, for the sake of contradiction, that adding a new validation activity $u$ to $U$ resolves the inconsistency. Let $U' = U \cup \{u\}$ be the new set of validation activities. For the set of validation activities to be consistent (Definition A7, $U'$ must be conjointly satisfiable, i.e., there exists an epistemic structure $M$ and a world $w$ such that $(M, w) \vDash U'$. Since $U' = U \cup \{u\}$, and $(M, w) \vDash U'$, it follows that $(M, w) \vDash U$. However, this contradicts the initial premise that $U$ is inconsistent. By the definition of *consistency* (definition A7), if $U$ is inconsistent, there cannot exist any epistemic structure $M$ and world $w$ such that $(M, w) \vDash U$. This contradiction shows that the earlier assumption that the addition of a new validation activity $u$ to $U$ resolves the inconsistency must be false.

**Theorem 35: *If two sets of valid validation activities $U_1$ and $U_2$ generate equivalent knowledge about the validation evidence for all validation criteria in $\Psi$, then these sets can be used interchangeably without affecting the validation of $\Psi$.***
<u>Proof:</u> Let $U_1$ and $U_2$ be two sets of valid validation activities, and $\Psi$ be the set of validation criteria. Assume $U_1$ and $U_2$ generate equivalent knowledge about $\Psi_e$, that is, for any epistemic structure M and world $w, (M, w) \vDash K_{U_1}(\Psi_e)$ iff $(M, w) \vDash K_{U_3}(\Psi_e)$. By the sufficiency condition in Definition 30, we have $K_{U_1}(\Psi_e) \vdash B(\Psi)$ and $K_{U_2}(\Psi_e) \vdash B(\Psi)$. Given the equivalence of knowledge generated by $U_1$ and $U_2$, we can conclude that they lead to the same beliefs about $\Psi$. The same argument holds for $\neg\Psi_e$ and $\neg B(\Psi)$. Therefore, $U_1$ and $U_2$ can be substituted for each other without affecting the validation of $\Psi$.

***Theorem 36:*** *If a set of validation activities **U** is valid, then the beliefs formed about the validation criteria $\Psi$ are necessarily derived from the knowledge generated by the validation evidence produced by **U**.*
*Proof:* Let U be a valid set of validation activities and $\Psi$ be the set of validation criteria. By the sufficiency condition in Definition 30, we have $K_U(\Psi_e) \vdash B(\Psi)$. This directly states that the knowledge generated by U about the validation evidence $\Psi_e$ leads to the belief in $\Psi$. Moreover, by the necessity condition, we have $\neg K_U(\Psi_e) \vdash \neg B(\Psi)$. This means that the absence of knowledge about the validation evidence necessarily leads to the absence of belief in the validation criteria. Together, these conditions demonstrate that the beliefs about the validation criteria $\Psi$ are necessarily derived from the knowledge generated by U. The sufficiency condition ensures that the knowledge is adequate to form the beliefs, while the necessity condition ensures that these beliefs cannot be formed without the knowledge generated by U. Therefore, we can conclude that if U is a valid set of validation activities, the beliefs formed about $\Psi$ are necessarily derived from the knowledge generated by the validation evidence produced by U.

***Theorem 37:*** *If two validation criteria are logically independent, then the order in which they are validated does not affect the overall validation assessment.*
*Proof:* Proof similar to Theorem 15.

***Definition 31 (Minimal Validation Activity Set):***
*A set of validation activities U is minimal with respect to validation evidence $\Psi_e$ if and only if:*
  a) **Validity:** *U is valid per Definition 30 (Valid Validation Activities)*
  b) **Essentiality (Non-Redundant):** *For every validation activity $u \in U$, there exists a validation criterion $\psi \in \Psi$ such that:*
     - The execution of u leads to knowledge about validation evidence $\psi_e$
     - The belief in this evidence entails belief in the criterion
     - Removing u breaks this relationship, i.e., $\forall u \in U, \exists \psi \in \Psi$ such that: $([u]K(\psi_e) \wedge B(\psi_e) \vDash B(\psi))$ but $\neg \exists u_2 \in (U \setminus \{u\})$ such that $[u_2]K(\psi_e) \wedge B(\psi_e) \vDash B(\psi)$

***Theorem 38:*** *Every valid set of validation activities contains at least one minimal set as a subset.*
*Proof:* This proof is similar in structure to Theorem 21.

***Theorem 39:*** *For a set of validation activities, the presence of a minimal set as a subset does not guarantee the validity of the containing set.*
*Proof:* This proof is similar in structure to Theorem 22.

***Theorem 40:*** *If a validation activity u in **U** is not essential according to our definition, then its removal does not affect the validation of any criteria.*
*Proof:* This proof is similar in structure to Theorem 23.

***Definition 32 (Validation assessment):***
*A validation Assessment, denoted as $A^{val}_{\psi_e}$, is the agent's belief about the value of the validation criterion $\psi$ based on the observed state of the validation evidence $\psi_e$, which can take on multiple values $\{\psi_{e1}, \psi_{e2}, \ldots, \psi_{en}\}$. Formally: $A^{val}_{\psi_e} := \bigwedge_{i=1}^{n} B(\psi_{ei} \rightarrow (\psi = x_i))$, where, B is the belief operator; $\psi_e$ is the validation evidence; $\psi_{ei}$ is the $i-th$ state of $\psi_e$; $\psi = x_i$ represents the agent assigning a specific value $x_i$ to the validation criterion $\psi$ when $\psi_e$ is observed to be in state $\psi_{ei}$; $x_i$ are specific values (e.g., numerical values, qualitative assessments) corresponding to each state $\psi_{ei}$.*

*Definition 33 (Validation confidence):*
Validation Confidence, denoted as $C_{A_{\psi_e}^{val}}$, is the degree of belief held by an agent about the validation assessment $A_{\psi_e}^{val}$. It represents the agent's confidence in their assessment of the validation criterion $\psi$ based on the observed levels of validation evidence $\psi_e$, which can take on multiple states. Formally: For each possible state $\psi_{ei}$ of the validation evidence $\psi_e$ and the corresponding assessment value $x_i$: $C_{A_{\psi_e}^{val}} := B((\psi_{ei} \rightarrow (\psi = x_i)) \rightarrow c_{xi}) \wedge B(\neg(\psi_{ei} \rightarrow (\psi = x_i)) \rightarrow c_{yi})$, where, $\psi_e$ is the validation evidence; $\psi_{ei}$ is the i-th state of $\psi_e$; $\psi = x_i$ represents the agent assigning the value $x_i$ to the validation criterion $\psi$ when $\psi_e$ is observed to be in state $\psi_{ei}$; $c_{xi}$ represents the proposition "The confidence in the assessment is $x_i$."; $c_{yi}$ represents the proposition "The confidence in the negation of the assessment is $y_i$."; $x_i$ and $y_i$ are numbers between 0 and 1, with $x_i + y_i = 1$.

This definition captures the agent's confidence in their validation assessment by associating confidence levels with each possible state of the validation evidence. The expression $B((\psi_{ei} \rightarrow (\psi = x_i)) \rightarrow c_{xi})$ represents the agent's belief that if the validation evidence is in state $\psi_{ei}$ and they assign $\psi = x_i$, then their confidence in this assessment is $x_i$. Conversely, $B(\neg(\psi_{ei} \rightarrow (\psi = x_i)) \rightarrow c_{yi})$ represents the agent's belief that if the evidence does not support the assessment $\psi = x_i$, then their confidence in the negation of the assessment is $y_i$.

*Definition 34 (Valid validation confidence):*
A validation confidence, $C_{A_{\psi_e}^{val}}$, is considered valid if it satisfies the following conditions for each possible state $\psi_{ei}$ of the validation evidence $\psi_e$:

- The agent knows the observed state of the validation evidence: $K(\psi_{ei})$
- **Uniqueness and Completeness:** For each validation assessment corresponding to $\psi_{ei}$, there exists exactly one pair of values $(x_i, y_i)$ such that $x_i + y_i = 1$, where, $c_{xi}$ represents the proposition "The confidence in the assessment is $x_i$.", and $c_{yi}$ represents the proposition "The confidence in the negation of the assessment is $y_i$."
- **Certainty:** The validation confidence is 1 (i.e., $x_i = 1, y_i = 0$) if and only if: $B(\psi_{ei} \Leftrightarrow (\psi = x_i)) \wedge K(\psi_{ei})$
- **Impossibility:** The validation confidence is 0 (i.e., $x_i = 0, y_i = 1$) if and only if: $B(\psi_{ei} \rightarrow (\psi \neq x_i)) \wedge K(\psi_{ei})$
- **Range**: The validation confidence is between 0 and 1 (i.e., $0 < x_i < 1, 0 < y_i < 1$) if and only if: $B(\psi_{ei} \rightarrow (\psi = x_i)) \wedge K(\psi_{ei}) \wedge \neg B(\psi_{ei} \Leftrightarrow (\psi = x_i))$
- **Non-Contradictory:** The agent cannot simultaneously believe that $\psi_{ei}$ implies $\psi = x_i$ and that $\psi_{ei}$ implies $\psi \neq x_i$: $\neg(B(\psi_{ei} \rightarrow (\psi = x_i)) \wedge B(\psi_{ei} \rightarrow (\psi \neq x_i)))$.

*Definition 35 (Proxy-Valid System Solution):*
A system solution, SysSol, is considered proxy-valid if and only if,
  a) For all stakeholder needs $\eta \in N$, there exists a set of validation criteria $\Psi$ where:
   - $B(\Psi) \vDash B(\eta)$ holds (sufficiency)
   - $\neg B(\Psi) \vDash \neg B(\eta)$ holds (necessity)
  b) Let $A^{val}$ be the set of validation assessments and $C_{A_{\psi_{ei}}^{val}}$ be the set of validation confidences. For each validation criterion $\psi \in \Psi$ and its associated validation evidence $\psi_e$, the system is proxy-valid if and only if: $\forall A_{\psi_{ei}}^{val} \in A^{val}: (C_{A_{\psi_{ei}}^{val}} \geq \tau)$
  c) Knowledge of validation evidence implies belief in the validation criterion: $K(\psi_e) \rightarrow B(\psi)$

This remains a "proxy" validity as the system solution is valid with respect to stakeholder needs, but not necessarily with respect to the actual stakeholder goals.

*Definition 36 (Valid System Solution):*
*A system solution, SysSol, is valid if and only if it produces outcomes that satisfy all stakeholder goals.* Let G be the set of all stakeholder goals, i.e., the real needs; O be the set of outcomes produced by the system solution; and S be the set of stakeholders. Then *SysSol* is goal-valid if and only if: $\forall g \in G: \exists o \in O: o \models g$, *where* $o \models g$ *denotes that outcome o satisfies goal g.*

***Theorem 41: Adding new validation evidence to an existing set of evidence cannot decrease the confidence in any validation criterion, as long as the new evidence is consistent with the existing evidence.***
*Proof:* Proof similar to Theorem 25.

***Theorem 42: A proxy-valid system solution cannot be a valid system solution if the needs themselves are invalid with respect to the stakeholder goals.***
*Formally, given a set of stakeholder goals G and an invalid set of needs N (i.e., N does not satisfy Definition 8's validity conditions), there exists a system solution SysSol where:* $(\forall \eta \in N: SysSol \models \eta) \land \exists g \in G: \forall o \in O: o \not\models g$.
*Proof:* Because N is invalid for G, believing N does not entail believing G, that is, either sufficiency or necessity for $N \rightarrow G$ fails (Definition 8). Let *SysSol* be the system solution that meets all needs in N. Since N is invalid for G, there must be at least one goal $g \in G$ that does not follow from satisfying N. Concretely, no outcome o of *SysSol* satisfies g. Hence *SysSol* is proxy-valid (it satisfies every $\eta \in N$) yet not goal-valid (it fails some goal *g*), all due to N itself being invalid with respect to G.

## XI. VERIFICATION VS VALIDATION

We use the theory elaborated in the previous section to investigate the relationship between verification and validation activities. Particularly, we inquire if, for a given problem and system solution, verification and validation activities are mutually exclusive, if there may be overlaps between them, or if one is a subset of the other. We investigate this aspect because of the lack of agreement in practice in this respect, as the traditional informal definitions are found unable, at least in practice, to provide clarity in some cases.

***Theorem 43: Let N be a set of needs, let θ be a requirement that is valid (sufficient and necessary) with respect to N. Suppose v is a valid verification activity for θ, then performing the verification activity v also induces belief in N.***
*Proof:* Given that $v$ is a valid verification activity for $\theta$, we know from Definition 19 that performing $v$ is sufficient to entail belief in $\theta$, which we can express formally as $[v]K(\varphi_e) \models B(\theta)$. Since $\theta$ is valid with respect to N, we know from Definition 12 that $\theta$ satisfies the sufficiency condition $B(\theta) \models B(N)$, telling us that belief in $\theta$ being satisfied entails belief in N being satisfied. Through transitivity of the entailment, since $[v]K(\varphi_e) \models B(\theta)$ and $B(\theta) \models B(N)$, we can conclude that $[v]K(\varphi_e) \models B(N)$. This means performing the verification activity $v$ leads to belief in N's satisfaction. Furthermore, by Definition 28, since $v$ produces evidence that builds belief in N's satisfaction, $v$ also functions as a validation activity for N.

***Theorem 44: Let N be a set of needs, let θ be a requirement that is neither necessary nor sufficient for N. Suppose v is a valid verification activity for θ, then performing the verification activity v does not induce a belief in N.***
*Proof:* When $\theta$ is neither necessary nor sufficient for N, we know by definition that both $B(\theta) \not\models B(N)$ and $\neg B(\theta) \not\models \neg B(N)$ hold. Even though $v$ is a valid verification activity, i.e., $[v]K(\varphi_e) \models B(\theta)$, there is no logical pathway to establish $[v]K(\varphi_e) \models B(N)$. The absence of both necessity and sufficiency conditions means there is no logical connection between beliefs about $\theta$ and beliefs about N.

*Theorem 45: Let N be a set of needs, let θ be a requirement that is sufficient but not necessary for N. Suppose v is a valid verification activity for θ, then performing the verification activity v induces belief in N.*

*Proof:* When $\theta$ is sufficient but not necessary for N, we know that $B(\theta) \vDash B(N)$ holds but $\neg B(\theta) \nvDash \neg B(N)$. Given that $v$ is a valid verification activity, we know $[v]K(\varphi_e) \vDash B(\theta)$. Through transitivity with the sufficiency condition, we can establish that $[v]K(\varphi_e) \vDash B(\theta)$ and $B(\theta) \vDash B(N)$ implies $[v]K(\varphi_e) \vDash B(N)$. The fact that $\theta$ is not necessary for N does not affect this chain of implications because we are only using the sufficiency direction of the relationship.

*Theorem 46: Let N be a set of needs, let θ be a requirement that is necessary but not sufficient for N. Suppose v is a valid verification activity for θ, then performing the verification activity v does not induce a belief in N.*

*Proof:* When $\theta$ is necessary but not sufficient for N, we know that $\neg B(\theta) \vDash \neg B(N)$ holds but $B(\theta) \nvDash B(N)$. Even though $v$ is a valid verification activity, meaning $[v]K(\varphi_e) \vDash B(\theta)$, we cannot establish $[v]K(\varphi_e) \vDash B(N)$ because $B(\theta) \nvDash B(N)$.

*Theorem 47: Let N be a set of needs and θ be a requirement. If a verification activity v is invalid, then performing v cannot induce beliefs about either θ or N, regardless of the validity relationship between θ and N.*

*Proof:* From Definition 19, we know an invalid verification activity means either $[v]K(\varphi_e) \nvDash B(\theta)$ or $\neg[v]K(\varphi_e) \nvDash \neg B(\theta)$. This means we cannot establish belief in $\theta$ through the activity, i.e., $[v]K(\varphi_e) \nvDash B(\theta)$. Without being able to establish $B(\theta)$, there is no way to establish $B(N)$ through any validity relationship between $\theta$ and N, as all such relationships would require $B(\theta)$ as a starting point.

*Theorem 48: Let Θ be a set of requirements and N be a set of needs. If Θ is sufficient but not necessary for N, then performing a valid set of validation activities for N does not necessarily induce belief in the satisfaction of Θ.*

*Proof:* When $\Theta$ is sufficient but not necessary for N, we know $B(\Theta) \vDash B(N)$ holds but $\neg B(\Theta) \nvDash \neg B(N)$. By contrapositive, this means $B(N) \nvDash B(\Theta)$. Given that $u$ is a valid validation activity, we know $[u]K(\psi_e) \vDash B(N)$. However, since $B(N) \nvDash B(\Theta)$, we cannot establish $[u]K(\psi_e) \vDash B(\Theta)$. Therefore, when $\Theta$ is only sufficient (and not necessary) for N, performing validation activities for N cannot support beliefs in $\Theta$'s satisfaction.

*Theorem 49: Let Θ be a set of requirements and N be a set of needs. If Θ is necessary but not sufficient for N, then performing a valid set of validation activities for N induces belief in the satisfaction of Θ.*

*Proof:* When $\Theta$ is necessary for N, we know $\neg B(\Theta) \vDash \neg B(N)$ holds. By contrapositive reasoning, this means $B(N) \vDash B(\Theta)$. Given that $u$ is a valid validation activity, we know $[u]K(\psi_e) \vDash B(N)$. Through transitivity with the relationship derived from necessity, since $[u]K(\psi_e) \vDash B(N)$ and $B(N) \vDash B(\Theta)$, we can establish $[u]K(\psi_e) \vDash B(\Theta)$. The fact that $\Theta$ is not sufficient for N (i.e., $B(\Theta) \nvDash B(N)$) does not affect this proof as we only require the necessity-derived relationship.

*Theorem 50: Let Θ be a set of requirements and N be a set of needs. If a validation activity u is invalid, then performing u cannot induce beliefs about either Θ or N, regardless of the validity relationship between Θ and N.*

*Proof:* This is similar in structure to Theorem 47.

These theorems help understand the formal relationship between verification and validation activities, as follows:

- If the requirements derivation process is believed to be correct (from needs to requirements), then a verification activity can serve as a validation activity.
- If the requirements derivation process is believed to be correct (from needs to requirements) and the operational solution is related to the requirements, then a validation activity can serve as a verification activity.
- If the operational solution is not related to the requirements, then regardless of the correctness of the derivation process, a validation activity cannot serve as a verification activity.

Therefore, the strength of the belief on the correctness of the requirements derivation activity is crucial to distinguishing verification from validation activities, or, more precisely, their usage. Furthermore, we can state in general that verification activities form a subset of validation activities, under the assumption of correct derivation, but we cannot guarantee that validation activities are a subset of verification activities.

## XII. CONCLUSION

This paper has presented a formal theory of verification and validation (V&V) in systems engineering, underpinned by Dynamic Epistemic Modal Logic (DEML). The work aims to resolve longstanding ambiguities in V&V by treating them as knowledge-building processes. To develop this theoretical foundation, we systematically formalize each component of the V&V framework, starting with the fundamental elements that V&V activities assess. This work formalized the distinction between needs and requirements, viewing them as representations of desired outcomes and required functions, respectively. This conceptual framing reveals how problem and solution spaces intersect: needs represent stakeholder goals, while requirements define the transformations needed by a system solution. By expressing these elements through DEML, the paper establishes how verification and validation activities build knowledge and beliefs about system properties through evidence.

Building on this framework, the paper introduced a DEML-based theoretical foundation that captures how evidence from V&V activities translates into beliefs about needs and requirements satisfaction. Through formal definitions and theorems, we establish precise conditions for validity, consistency, sufficiency, necessity, and minimal sets. These theorems prove when needs, requirements, verification criteria and validation criteria support belief formation about system properties. A key insight is that inconsistency at any level fundamentally breaks the chain of belief formation that V&V activities aim to establish.

The paper then applies this formal foundation to analyze how verification and validation activities inform beliefs about requirements and needs satisfaction. Through a series of theorems, we prove the precise conditions under which verification activities that build belief in requirements also inform beliefs about needs satisfaction, and vice versa. When requirements are both necessary and sufficient for needs, verification naturally informs validation. When requirements are only sufficient but not necessary, verification can inform beliefs about needs satisfaction but validation may not inform beliefs about requirements satisfaction. When requirements are necessary but not sufficient, validation activities inform beliefs about requirements satisfaction even though verification cannot inform beliefs about needs satisfaction. These results provide a rigorous logical foundation for understanding how V&V activities build knowledge about systems.

Beyond these core theoretical findings, the paper has broader methodological implications. The DEML-based perspective paves the way for model-checkers or advanced reasoning engines capable of automatically detecting inconsistent requirements or insufficient verification evidence. By making explicit the logical relationships between needs, requirements, verification criteria, and validation criteria, teams can more transparently communicate the rationale behind each V&V decision.

The theory moves V&V from informal practices to a framework grounded in formal logic. By proving exactly when verification activities inform beliefs about needs satisfaction and when validation activities inform beliefs about requirements satisfaction, it improves traceability, reduces missed contradictions, and fosters greater confidence in engineering outcomes. These benefits suggest promising avenues for future work, such as extending the logic to partial evidence, exploring minimal verification sets in large-scale systems, or integrating these methods into model-based systems engineering tools.

**APPENDIX: BACKGROUND ON DYNAMIC EPISTEMIC MODAL LOGIC**

**A. Syntax for dynamic epistemic logic:**

The syntax for dynamic epistemic logic consists of a non-empty set ($\Phi$) of atomic propositions that represent the statements about verification evidence, and requirement statements about system characteristics. Atomic propositions may be combined using conjunction and/or negation to form compound sentences or formulas, which are usually denoted by Greek symbols $\varphi, \psi$, etc. In addition, the formula $\varphi$ can be replaced with simple propositions $p$, and/or compounded formulas with the knowledge operator *(K)*, the belief operator *(B)*, and the event *([V])* operator, and/or a combination of these. Here $K(\varphi)$ means that "the agent knows that $\varphi$", $B(\varphi)$ means that "the agent believes that $\varphi$", and $[V](\varphi)$ means that "after the event *V* has taken place, $\varphi$ is true". The knowledge and belief operator can be combined with the event operator to represent verification and validation activities (see Definitions 17 and 28). As an example, $[v]K(\varphi_e \vee \neg\varphi_e)$ means after performing the event (i.e., the verification activity – $v$) the agent knows the truth or falsity of the verification evidence, $\varphi_e$.

*Definition A1 (Dynamic epistemic language):*
*The set of atomic propositions, formulas, connectives and the modal operators $K, B,$ and $[D]$, along with the agent number (denoted by the subscript), collectively form a formal language $L_d$, called dynamic epistemic language ($L_d$), which is given by the following Backus-Naur/Backus-Normal Form (BNF) [ref].*

$$\varphi := p \mid \neg\varphi \mid (\varphi \wedge \psi) \mid K(\varphi) \mid B(\varphi) \mid [D](\varphi)$$

BNF is a formal notation representing the grammar of the formal language. := means "may be expanded to" or "replaced with."

## B. Semantics for dynamic epistemic logic:

*Definition A2 (Epistemic structure):*
*An epistemic structure, similar to Kripke structure [ref], is a tuple $M_i = (S_i, P_{K_i}, P_{B_i}, \pi_i)$, where the subscript $i$ represents the agent $i$. $S_i$ is the set of all worlds, also sometimes called the domain of $M_i$, and $P_{K_i}$ and $P_{B_i}$ are binary relations, called accessibility relations, that are used to evaluate knowledge and belief respectively. The epistemic structure contains all the necessary elements required to represent knowledge and belief as modals. The $\pi_i$ in the structure $M_i$ is a valuation function, as defined in definition A4.*

*Definition A3 (Accessibility relation):*
*The accessibility relations ($P_{K_i}$ and $P_{B_i}$) in the epistemic structure (definition A2) are binary relations on $S$ and they represent the set of worlds agent $i$ considers possible at a particular world. Mathematically, both $P_{K_i}$ and $P_{B_i}$ can be defined using Eq. (1). However, the properties of these relations change with respect to knowledge or beliefs.*

$$\begin{aligned}\mathcal{P}_{K_i}(w) &= \{w' : (w, w') \in \mathcal{P}_{K_i}\} \\ \mathcal{P}_{K_i} &\in S \times S\end{aligned} \quad (1)$$

This can be visualized using a directed graph as shown in Fig. 1.

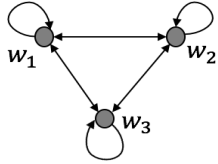

**Figure 1**. Possible worlds.

The arrows indicate worlds that an agent considers possible at a particular world. For example, at $w_1$, the agent considers $w_1$, $w_2$ and $w_3$ to be the possible worlds. More generally, ($w,w'$) belongs to $\mathcal{P}_{K_i}$ if the agent considers $w'$ a possible world in world $w$. The accessibility relation can be defined appropriately to reflect the properties of knowledge or belief. For instance, while representing what an agent *knows*, $\mathcal{P}_{K_i}$ is considered to be an equivalence relation, i.e., reflexive, transitive, and symmetrical. Considering the accessibility relation to be reflexive implies that the agent always considers the actual state ($w$) to be in the set of possible states ($P_{K_i}$). This is the primary property that distinguishes knowledge from belief. The accessibility relation for belief ($P_{B_i}$) is considered to be serial, Euclidean, and transitive. The structure of accessibility relation leads to some axioms and rules of inferences that define an axiomatic system for knowledge and belief.

*Definition A4 (Valuation function):*
*In the epistemic structure $M_i = (S_i, P_{K_i}, P_{B_i}, \pi_i)$, $\pi$ is a valuation function, that assigns truth values to each of the atomic propositions in $\Phi$ (the set of all atomic propositions) at each state, i.e., $\pi(w,p)$ = TRUE means that the proposition p is true at state w. The state w is emphasized here as the truth assignment changes when the state changes.*

$$\pi(w): \Phi \rightarrow \{TRUE, FALSE\} \; for \; each \; state \; w \in S \tag{2}$$

With the elements of the epistemic structure defined, the semantic relation can be recursively defined as $(M, w) \vDash \varphi$ which can be read as "$\varphi$ is true in structure $M$ at state $w$" or "structure $M$ satisfies $\varphi$ at state $w$". Equation (2) states that atomic proposition $p$ is TRUE at state $w$ in structure $M$, if and only if $\pi$ assigns *TRUE*. This is the same as in propositional logic.

$$(M, w) \vDash p \; iff \; \pi(w, p) = TRUE \tag{3}$$

Also, we have

$$(M, w) \vDash \varphi \wedge \psi \; iff \; (M, w) \vDash \varphi \; and \; (M, w) \vDash \psi \tag{4}$$
$$(M, w) \vDash \neg \varphi \; iff \; (M, w) \nvDash \varphi$$

Similarly, "$(M, w)$ satisfies the propositions that agent *i knows $\varphi$*", and "$(M, w)$ satisfies the propositions that agent *i believes $\varphi$*", are defined by Eqs. (5) and (6) respectively:

$$(M, w) \vDash K(\varphi) \Leftrightarrow (M, w') \vDash \varphi \; for \; all \; w' \in P_{K_i}(w) \tag{5}$$
$$(M, w) \vDash B(\varphi) \Leftrightarrow (M, w') \vDash \varphi \; for \; all \; w' \in P_{B_i}(w) \tag{6}$$
$$(M, w) \vDash [D](\varphi) \Leftrightarrow (M, w) \vDash [D] \rightarrow \varphi \tag{7}$$
$$(M, w) \vDash [D]\big(K(\varphi)\big) \Leftrightarrow (M, w) \vDash [D] \rightarrow K(\varphi) \tag{8}$$
$$(M, w) \vDash [D]\big(B(\varphi)\big) \Leftrightarrow (M, w) \vDash [D] \rightarrow B(\varphi) \tag{9}$$

Equation (5) means that the agent *knows* that $\varphi$ if and only if $\varphi$ is true in all the worlds the agent considers possible at world $w$ in structure $M$, and the possible worlds are characterized by the accessibility relation for knowledge ($P_{K_i}$). Equation (6), on the other hand, means that the agent *believes* that $\varphi$ if and only if $\varphi$ is true in all the worlds the agent considers possible at world $w$ in structure $M$, and the possible worlds are characterized by the accessibility relation for belief ($P_{B_i}$). The properties of the notion of knowledge and belief are characterized through axioms and rules of inference that represent some form of idealizations [31] as seen in Table I.

In Dynamic Epistemic Modal Logic (DEML), dynamic operators are used to represent how events or actions change the epistemic state of agents, affecting their knowledge and beliefs. This paper introduces a simplified semantics of these operators to make the concept more accessible in the context of V&V processes. In Eq. 7, $(M, w) \vDash [D](\varphi) \Leftrightarrow (M, w) \vDash [D] \rightarrow \varphi$, means that after the event [D], the statement $\varphi$ is true. Similarly in Eq. 8, $(M, w) \vDash [D]\big(K(\varphi)\big) \Leftrightarrow (M, w) \vDash [D] \rightarrow K(\varphi)$, means that the agent knows $\varphi$ after the occurrence of [D]. Finally in Eq. 9, $(M, w) \vDash [D]\big(B(\varphi)\big) \Leftrightarrow (M, w) \vDash [D] \rightarrow B(\varphi)$, means that the agent believes $\varphi$ after the occurrence of [D]. While this simplified semantics serves to introduce the concept of dynamic change in epistemic states, it is important to note that the full DEML framework employs more sophisticated mechanisms to model these changes. Specifically, DEML uses action models and product updates to provide a comprehensive account of how events transform agents' knowledge and beliefs [26, 27, 32].

Axiomatic logic systems can be formed by choosing axioms and rules of inference to reflect the problem of interest. The validity of these axioms depends on how one represents the accessibility relation. For example, if we consider $P_{K_i}$ to be an equivalence relation (reflexive, symmetric and transitive), then axiom T results from $P_{K_i}$ being reflexive, axiom 4 from transitive and axiom 5 from symmetric and transitive. We can similarly create various axiomatic systems to capture what we want the accessibility relation to look like. For instance, system **S5** is typically used to capture properties of knowledge. **S5** includes axioms K, T, 4 and 5. System **KD45** is used in the case of beliefs. The truth axiom (T) in **S5** is replaced by the consistency axiom (D) in **KD45**. This follows from the definition of the accessibility relation for belief ($P_{B_i}$) to be serial, Euclidean, and transitive. As can be seen in Table I, the truth axiom means that the agent can only know statements that are objectively true, which is not required for belief. In this paper, we will use system **KD45** to represent and reason about beliefs of agents.

**Table I. Axioms and Rules of Inferences**

| Axioms/Rules of inference | Mathematical representation | Description |
|---|---|---|
| All instances of propositional tautologies | | All axioms and rules of inference associated with Propositional Logic. |
| Distribution axiom (K) | $K_i\varphi \wedge K_i(\varphi \Rightarrow \psi)) \Rightarrow K_i\psi$ | Agents are powerful reasoners who know all the logical consequences of their knowledge. For example, if an agent *knows* $\varphi$ and *knows* that $\varphi$ *implies* $\psi$, then he *knows* $\psi$. |
| Knowledge generalization rule (N) | $From\ \varphi\ infer\ K_i\varphi$ | This means that if a statement $\varphi$ is true in all the states the agent considers possible, then the agent *knows* $\varphi$. This does not mean that the agent knows all statements that are true. |
| Knowledge or Truth axiom (T) | $K_i\varphi \Rightarrow \varphi$ | Agent only *knows* things that are true. |
| Positive introspection axiom (4) | $K_i\varphi \Rightarrow K_iK_i\varphi$ | Axioms 4 and 5 mean that the agent has introspection of his own knowledge base, i.e., the agent knows what he knows and does not know. |
| Negative introspection axiom (5) | $\neg K_i\varphi \Rightarrow K_i\neg K_i\varphi$ | |
| Consistency axiom (D) | $\neg B_i \bot$ | Axiom D means that the agent does not believe in a statement and its negation, i.e., $B_i\varphi \Rightarrow \neg B_i\neg\varphi$ |
| Modus ponens | $From\ \varphi\ and\ \varphi \Rightarrow \psi\ infer\ \psi$ | |

### C. Representing Conditional Beliefs in DEML

While the Dynamic Epistemic Modal Logic framework presented in this paper does not directly support the representation of conditional probabilities, we can express similar concepts using the existing syntax and semantics of DEML. This section explains how conditional beliefs can be represented within the constraints of DEML.

In probability theory, a conditional probability $P(A|B)$ represents the probability of event A occurring given that event B has occurred. In the context of V&V, we might want to express beliefs about a system's properties given certain evidence. However, DEML does not have a direct equivalent to the "|" operator used in conditional probabilities. Instead, we can represent conditional beliefs in DEML using a combination of the belief operator B and material implication. Consider the following example: Let $q$ represent the proposition "Temperature sensors show readings between 20°C and 25°C" (verification evidence). Let $p$ represent the proposition "The system performs within specified temperature range" (verification criterion). To express the belief that if the sensors show readings in the specified range, then the system performs within its specified range, we can write: $B(q \rightarrow p)$. This formula reads as "The agent believes that if q is true, then p is true." To capture a more complete picture of the agent's beliefs, including what they believe when the evidence is absent, we can use a conjunction of two belief statements: $B(q \rightarrow p) \wedge B(\neg q \rightarrow \neg p)$. This expanded formula represents: The agent believes that if the sensors show readings in the specified range, the system performs within its specified range. The agent also believes that if the sensors do not show readings in the specified range, the system does not perform within its specified range. The representation of conditional beliefs as described above will be used throughout this paper when discussing V&V assessments, criteria, and related concepts.

*Definition A5 (Logical validity):*
*In formal logic, an argument is logically valid if and only if it takes a form that makes it impossible for the premises to be true and the conclusion to be false. In the example below, the argument is valid since if both the premises are true, then the conclusion is true.*

*Premise 1*: If Tom is a Martian, then he was born on Mars
*Premise 2*: Tom is a Martian
*Conclusion*: Tom was born on Mars

*Validity* of arguments is defined structurally and is not concerned with the actual meaning of the sentences in the argument. Logicians typically tie logical consequence (logical entailment) to define *validity*. Specifically, there are two logical consequences: 1) Semantic or model-theoretic consequence, and 2) Syntactic or proof-theoretic consequence. With semantic consequence, an argument is *valid* if and only if there is no counter example. Mathematically, a semantic consequence can be represented as in Equation (7):

$$\Gamma \vDash A \tag{7}$$

Here $A$ is a semantic consequence of a set of premises $\Gamma$. "$A$" is a semantic consequence if and only if for every interpretation that makes $\Gamma$ true, it also makes $A$ true. Syntactic consequence, on the other hand, can be represented by Equation (8):

$$\Gamma \vdash A. \tag{8}$$

Here $A$ is a syntactic consequence of a set of premises $\Gamma$ if and only if $A$ can be derived from $\Gamma$ by the application of inference rules that are part of the formal system. Most logicians typically associate *validity* of arguments with semantic consequence, as syntactic consequence focuses more on the formal procedure associated with just symbol manipulation. In this paper, validity with respect to different SE artifacts (needs, requirements, system, etc.) will be defined using this notion of semantic consequence.

*Definition A6 (Satisfiability):*
*A formula $\varphi$ is satisfiable if there exists some structure $M$ and some state $w \in S$ for $M$ such that $(M, w) \vDash \varphi$. A set of formulas $\Phi$ is satisfiable if and only if there exists some structure $M$ and some state $w \in S$ for $M$ such that $\forall_{\phi \in \Phi}(M, w) \vDash \phi$.* Given a set of sentences, one can determine whether the set is consistent or not using this notion of satisfiability. If the set of sentences is inconsistent, then the set of sentences are not satisfiable.

*Definition A7 (Consistency):*
*A set of formulas $\Phi$ is* consistent *if and only if it is satisfiable (Definition A6). In other words, a set $\Phi$ is consistent if and only if there exists an epistemic structure $M$ and a world $w$ in $M$ such that $(M, w) \vDash \Phi$, where $(M, w) \vDash \Phi$ means that $(M, w) \vDash \varphi$ for all $\varphi \in \Phi$.*